%% file: FL.tex
\UseRawInputEncoding  %%for arxiv compile

\documentclass[times,twocolumn,final]{elsarticle}

\usepackage{lineno}
\usepackage{framed,multirow}
\usepackage{amssymb}
\usepackage{latexsym}

\usepackage{url}
\usepackage{xcolor}

\usepackage{graphicx}
\usepackage{amsmath}
\usepackage{amssymb}
\usepackage{verbatim}
\usepackage{booktabs}
\usepackage{bm}
\usepackage{array}
\usepackage{epstopdf}
\usepackage{makecell}
\usepackage{tabularx} % added
\usepackage{booktabs}   %three lines for table, added
\usepackage{mathrsfs} %added
\usepackage{framed, multirow} %added
\usepackage{amsfonts} %added
\usepackage{wrapfig}  %added
\usepackage{cite}
\usepackage{footnote}

\usepackage{latexsym}
\usepackage[caption = false]{subfig}
\usepackage{graphicx}

\usepackage{tablefootnote}
\usepackage{float}
\usepackage[hidelinks,colorlinks=true,linkcolor=blue,citecolor=blue]{hyperref}

\input{defs}
\usepackage{geometry}
\begin{document}
	
	\begin{frontmatter}
		\title{Deep learning for unsupervised domain adaptation in medical imaging: Recent advancements and future perspectives}

\author[mymainaddress]{Suruchi Kumari}\ead{suruchi\_k@cs.iitr.ac.in}
\author[mymainaddress]{Pravendra~Singh\corref{mycorrespondingauthor}}\ead{pravendra.singh@cs.iitr.ac.in}

\cortext[mycorrespondingauthor]{Corresponding author: Pravendra Singh }

\address[mymainaddress]{Department of Computer Science and Engineering, Indian Institute of Technology Roorkee, India}

\begin{abstract}
%%%
Deep learning has demonstrated remarkable performance across various tasks in medical imaging. However, these approaches primarily focus on supervised learning, assuming that the training and testing data are drawn from the same distribution. Unfortunately, this assumption may not always hold true in practice. To address these issues, unsupervised domain adaptation (UDA) techniques have been developed to transfer knowledge from a labeled domain to a related but unlabeled domain. In recent years, significant advancements have been made in UDA, resulting in a wide range of methodologies, including feature alignment, image translation, self-supervision, and disentangled representation methods, among others. In this paper, we provide a comprehensive literature review of recent deep UDA approaches in medical imaging from a technical perspective. Specifically, we categorize current UDA research in medical imaging into six groups and further divide them into finer subcategories based on the different tasks they perform. We also discuss the respective datasets used in the studies to assess the divergence between the different domains. Finally, we discuss emerging areas and provide insights and discussions on future research directions to conclude this survey. %%%%
\end{abstract}


\end{frontmatter}

%\linenumbers
%%%%%%%%%%%%%%%%%%%%%%%%%%%%%%%%%%%%%%%%%%%%%%%%%%%%%%%%%%%%%%%%%%%%%%%%%%%%%%%%%%%
%%%%%%%%%%%%%%%%%%%%%%%%%%%%%%%%%%%%%%%

\section{Introduction \label{Intro}}

Medical imaging informatics utilizes digital image processing and machine learning (ML) to improve the efficiency, accuracy, and reliability of imaging-based diagnosis \citep{mendelson2013imaging}. In recent years, significant progress has been made in medical imaging due to the increasing availability of data and the rapid development of deep learning (DL) techniques \citep{litjens2017survey}. However, the effective application of deep learning models in clinical contexts is hampered by fundamental difficulties. Annotated medical datasets are scarce due to the time-consuming labeling procedure \citep{litjens2017survey} and are difficult to share because of privacy issues \citep{adler2012sharing, sharma2019preserving}. Multicenter datasets can improve the availability of annotated data; however, the data is heterogeneous due to variations in hospital practices and patient groups \citep{zhang2020collaborative}. Domain shift occurs when there is a disparity between the distribution of the data used to train a deep learning model and the distribution of the data encountered during testing. DL models primarily rely on the oversimplified assumption that the source (training) and target (test) data are independent and identically distributed (i.i.d.) \citep{zhou2022domain}. However, when this assumption is violated, a classifier trained on the source domain is likely to exhibit reduced performance when tested on the target domain due to domain differences. Previous research has shown that domain shift between training and test datasets generally leads to an increase in test error \citep{ben2006analysis}. Therefore, understanding how to address domain shift is critical for effectively applying DL methods to medical image analysis.

\begin{figure*}
    \centering
    \subfloat[(i)]{\includegraphics[width = 9cm]{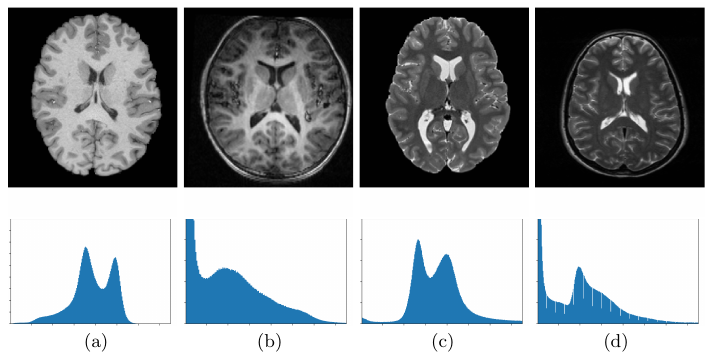}}
    \hspace{0.5cm}
    \subfloat[(ii)]{\includegraphics[width=8cm]{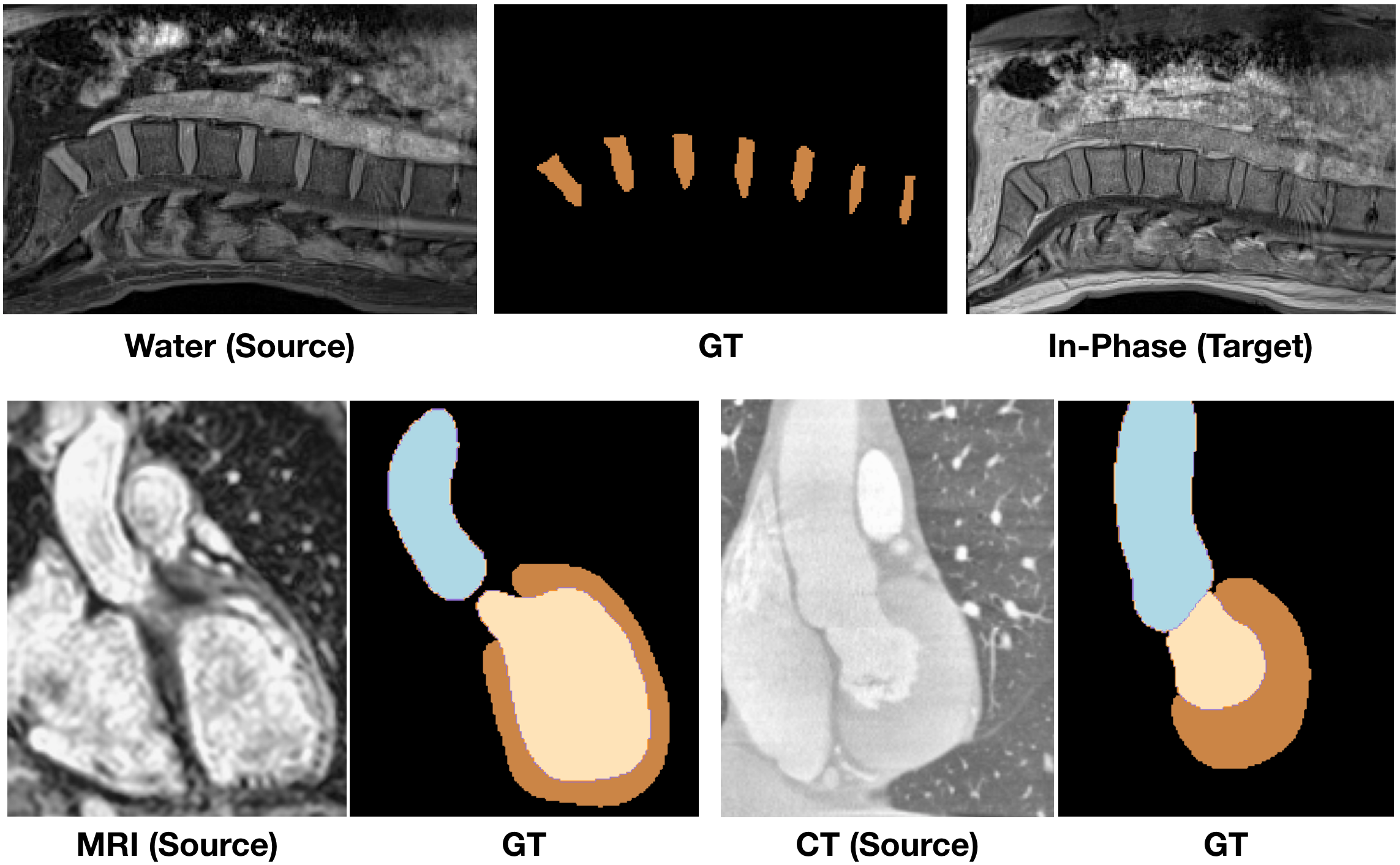}}
   
    \caption{Visualization of severe domain shifts between source and target modalities in two applications. From left to right (i) Image slices (top) and corresponding intensity distribution (bottom) of normalized T1-weighted (a, b) and T2-weighted (c, d) MRIs from different scanners. Image courtesy of Karani et al. \citep{karani2018lifelong}. (ii) Top: Two aligned spine images from Water and In-Phase MRI and the corresponding ground-truth segmentation, with the intervertebral disks depicted in brown and the background in black. Bottom: Two cardiac images from MRI and CT, and their ground-truth segmentations. The cardiac structures of AA, LVC, and MYO are depicted in blue, purple, and brown, respectively. Image courtesy of Bateson et al. \citep{bateson2021constrained}.}
    \label{fig_1}
\end{figure*}

The presence of domain shift is a common challenge in real-life applications, particularly in the field of biomedical image analysis. Biomedical radiological images are captured using various imaging modalities, such as Computerized Tomography (CT) and Magnetic Resonance (MR) imaging, which differ from natural images captured by optical devices. These modalities have distinct imaging physics principles, leading to significant differences in data distributions \citep{dou2018unsupervised}. The appearance and intensity histograms of anatomical structures vary across radiology modalities (Fig.~\ref{fig_1}). Domain adaptation (DA) has gained attention in the medical image analysis field as a means to address distribution discrepancies among related domains. Unsupervised domain adaptation (UDA) specifically addresses scenarios where labeled source data and only unlabeled target data are available for training \citep{guan2021domain}. UDA has become prominent in medical imaging due to its ability to adapt labeled data to new applications, reducing the need for expensive labeled data in the target domain. Several researchers have employed domain adaptation strategies to tackle diverse challenges in medical image analysis, including cancer tissue classification \citep{ren2018adversarial, zhang2019whole}, cross-modality segmentation \citep{wu2021unsupervised, cui2021structure}, nuclei instance segmentation \citep{liu2020unsupervised, liu2020pdam}, and cell detection \citep{xing2019adversarial, xing2020bidirectional}, among others.

Domain adaptation \citep{wang2018deep, wilson2020survey} and transfer learning \citep{kouw2018introduction} using real-world images have been extensively reviewed. However, the evaluation of domain adaptation and its applications in medical image analysis is relatively limited. Previous domain adaptation surveys lack in-depth coverage and comparison of unsupervised deep domain adaptation approaches. Guan and Liu \citep{guan2021domain} conducted a survey that examined shallow and deep learning-based domain adaptation methods, but since then, numerous research papers on the topic have emerged that were not covered in their study. Sarafraz et al. \citep{sarafraz2022domain} conducted a survey specifically on domain adaptation and domain generalization techniques in functional brain signals. Choudhary et al. \citep{choudhary2020advancing} reviewed domain adaptation methods for medical imaging, but it covers a relatively small number of techniques compared to our study.

In all the mentioned works, their focus was on domain adaptation, while our focus is specifically on deep unsupervised domain adaptation for medical imaging. In this study, we comprehensively examine and discuss current advancements and challenges in deep unsupervised domain adaptation for medical image analysis. The major contributions of our work are as follows:
\begin{itemize}
    \item This is the first survey paper that comprehensively covers recent advances in deep unsupervised domain adaptation for medical imaging. Specifically, we present a comprehensive overview of more than 140 relevant papers to cover the recent progress.
    \item We present an in-depth discussion of deep unsupervised domain adaptation methods, as depicted in Fig.~\ref{fig_3}. We categorize these methods into six groups and further divide them based on the specific tasks they perform, such as classification, segmentation, detection, medical image synthesis, depth estimation, and others.
    \item We also present a comprehensive analysis of different deep unsupervised domain adaptation methods in Table~\ref{table:Unsupervised}, emphasizing the specific characteristics of the datasets employed as the source and target domains in each study. This analysis aims to assess the divergence between the two domains in the respective works.
    \item Lastly, we explore emerging areas of deep unsupervised domain adaptation methods and highlight several potential future directions for further research and development in this domain.
\end{itemize}

The structure of this survey is outlined as follows. Section~\ref{sec_background} provides a brief introduction to the background and an overview of various learning schemes in domain adaptation. Section~\ref{sec_3} presents a comprehensive discussion on recent advances in deep unsupervised domain adaptation (UDA) methods in medical image analysis. Section~\ref{sec_4} explores emerging areas of deep UDA methods. In Section~\ref{sec_5}, potential research directions are discussed. Finally, Section~\ref{sec_6} concludes this survey paper.

%%%%%%%%%%%%%%%%%%%%%%%%%%%%%%%%%%%%%%%%%%%%%%%%%%%%%%%%%%%%%%%%%%%%%%%%%%%%%%%%%
\section{Background \label{sec_background}}

\subsection{Types of domain shift \label{subsec_2.1}}

A domain $D$ consists of a feature space $X$ and the task is defined by the label space $Y$ then the marginal distributions of $X$ and $Y$ in the source (S) and the target (T) domains are denoted by $P_s(X)$, $P_s(Y)$, $P_t(X)$, and $P_t(Y)$, respectively. Similarly, the conditional distributions in the two domains are denoted by $P_s(X|Y)$, $P_s(Y|X)$, $P_t(X|Y)$, and $P_t(Y|X)$. Domain shifts are of three types \citep{kouw2018introduction}. The first type is known as prior shift or the class imbalance problem. It occurs when the prior distributions of classes differ between domains $P_s(Y) \neq P_t(Y)$, but the conditional distributions are similar $P_s(X|Y) = P_t(X|Y)$. The second type is concept shift, also referred to as data drift. It occurs when the conditional distributions vary across domains $P_s(Y|X) \neq P_t(Y|X)$ while the data distributions remain constant $P_s(X) = P_t(X)$. The third type is covariate shift. It refers to a situation in which the marginal distribution of the input variables, also known as covariates or features, varies across different domains $P_s(X) \neq P_t(X)$, while the conditional distribution of the output variables given the inputs remains consistent $P_s(Y|X) = P_t(Y|X)$. In simpler terms, it means that the distribution of the input data is different between domains, but the relationship between the input and output variables remains the same.

The majority of the suggested domain adaptation approaches seek to address covariate shift. In medical imaging, covariate shift can arise from the usage of various imaging modalities (such as MRI and CT), variations in image acquisition equipment (such as different scanners within the same modality), or discrepancies in anatomical characteristics among the patient population (e.g., differences between males and females). Mathematically, domain adaptation (DA) can be defined as follows: Let $\mathcal{X} \times \mathcal{Y}$ represent the joint feature space and the corresponding label space, respectively. A source domain S and a target domain T are defined on $\mathcal{X} \times \mathcal{Y}$, with different distributions $P_s$ and $P_t$, respectively. Suppose we have $n_s$ labeled samples in the source domain, denoted as $\mathcal{D}_S = {(\mathbf{x}^S_i, y^S_i)}^{n_s}_{i=1}$. In the target domain, we have $n_t$ samples with  labels, $\mathcal{D}_T = {(\mathbf{x}^T_j, y^T_j)}^{n_t}_{j=1}$, or without labels, $\mathcal{D}_T = {(\mathbf{x}^T_j)}^{n_t}_{j=1}$. The goal of DA is to transfer the knowledge learned from the source domain (S) to the target domain (T) in order to perform a specific task on T. This task is shared by both the source and target domains. Unsupervised domain adaptation (UDA) specifically addresses scenarios where only unlabeled target data is available for training, in addition to labeled source data.

\begin{table*}[t]
    \centering
    \scriptsize
    \begin{center} 
  %  \tabstyle{5pt}
    \caption{Commonly used domain adaptation datasets.}
    \label{tab:datasets}
   \resizebox{\textwidth}{!}{
    \begin{tabular}{p{2.5cm}p{1.5cm}p{2cm}p{3cm}p{5cm}p{3.5cm}}
    \toprule
  Dataset & \multicolumn{1}{l}{Organ} & \multicolumn{1}{l}{Types} & \multicolumn{1}{l}{Task}  & \multicolumn{1}{l}{Description} & \multicolumn{1}{l}{Characterization of domain shift}\\ \midrule

BraTS \citep{menze2014multimodal} & Brain & MR images & Brain tumor segmentation & Manualy segmented: Annotations comprise the GD-enhancing tumor (ET-label 4), the peritumoral edema (ED-label 2), and the necrotic and non-enhancing tumor core (NCR/NET-label 1) & Acquired with different clinical protocols and various scanners from multiple (19) institutions, it consists of four different contrasts - T1, T1c, T2, and FLAIR  \\ \midrule

MICCAI WMH Challenge \citep{kuijf2019standardized}  & Brain & MR images & White matter hyperintensities (WMH) segmentation & WMH and other pathologies (i.e. lacunes and non-lacunar infarcts, (micro) hemorrhages) were manually segmented &Five different scanners in three different institutes  \\ \midrule

MM-WHS challenge dataset \citep{zhuang2016multi, zhuang2019evaluation} & Whole heart & MR and CT images & Whole heart segmentation  & Manually labeled: left and right ventricular cavity (LV, RV), left and right atrial cavity (LA, RA), myocardium of the left ventricle (Myo), ascending aorta trunk and pulmonary artery (PA) trunk & Cross-modality dataset \\ \midrule

REFUGE challenge dataset \citep{orlando2020refuge} & Eye & Fundus images & 1.Classification of clinical Glaucoma \newline 2.Segmentation of Optic Disc (OD) and Cup (OC) \newline 3.Localization of Fovea (macular center) & 1.Annotated as glaucoma and non-glaucoma images \newline 2.Manual pixel-wise annotations of the optic disc and cup \newline 3.Manual pixel-wise annotations of the fovea (macular center) & Acquired from different camera \\ \midrule

SCGM dataset \citep{prados2017spinal} & Spinal Cord & MRI images  & Spinal cord gray matter segmentation & Manual segmentation using different software packages of spinal cord grey matter (GM) and white matter (WM) tissue & Acquired from 4 medical centers \\ \midrule 

MICCAI2018 IVDM3Seg dataset & Intervertebral Disc & MRI images & Intervertebral discs (IVD) localization and segmentation & Manual segmentation for each IVD in the form of binary mask for lower spine & Different Modalities (in-phase, opposed-phase, fat and water images)  \\ \midrule

PROMISE12 challenge dataset \citep{litjens2014evaluation} & Prostate & MR images & Prostate segmentation & Annotations were performed on a slice-by-slice basis using a contouring tool for prostate capsule &  4 different centers with differences in scanner manufacturer, field strength and protocol \\ \midrule

CAMELYON17 dataset \citep{litjens20181399} & Breast & Whole-slide images (WSIs) & Detection and classification of breast cancer metastases & On a lesion-level: with detailed annotations of metastases in WSI and on a patient-level: with a pN-stage label per patient. & Acquired from five medical centers \\ \midrule

RIGA+ dataset \citep{hu2022domain} & Eye & Fundus images & Segmentation of Optic Disc (OD) and Cup (OC) & OC and OD boundaries of images were marked and annotated manually by ophthalmologists individually using a tablet and a precise pen. & Acquired from different sources. \\ \midrule

MR T1-weighted volumetric brain imaging dataset: Calgary-Campinas-359 (CC-359) \citep{souza2018open} & Brain & MR images & Skull stripping or Brain segmentation  & Consensus segmentation masks were generated for each subject using the Simultaneous Truth and Performance Level Estimation (STAPLE) method and manual segmentation is performed for twelve subjects & Acquired on scanners from three vendors (Siemens, Philips and General Electric) at both 1.5 T and 3 T magnetic field strengths \\ \midrule

ADNI dataset \citep{jack2008alzheimer, weiner2017alzheimer} & Brain & MRI, PET, fMRI, etc.. & Alzheimer’s Disease identification & Manualy annotated through a combination of clinical assessments, cognitive tests, and imaging data. & Cross-modality dataset \\ \midrule

The Cancer Genome Atlas (TCGA) dataset \citep{kandoth2013mutational} & Prostate & Histopathology WSIs & Cancer tumour classification based on gleason scores  & Gleason scores manually annotated by pathologists ranging from 6 to 10 & Acquired from 32 Clinical centers \\ \midrule

REST-meta-MDD Consortium \citep{yan2019reduced} & Brain & Resting-state functional magnetic resonance imaging (R-fMRI) & Major Depressive Disorder (MDD) classification & Images are annotated as MDD patients and healthy controls (HCs)  &
 Images acquired from 25 research groups from 17 Chinese hospitals/sites \\  \bottomrule

\end{tabular}
}
\end{center}

\end{table*}

\subsection{Datasets and applications \label{subsec_2.3}}

Domain adaptation has been extensively studied across various tasks in medical imaging, including segmentation, classification, detection, and others. In Table~\ref{tab:datasets}, we provide an overview of commonly used datasets that are specific to different tasks. In the following sections, we briefly discuss these tasks and their respective datasets.

\textbf{Image classification:}  
The goal of the classification model is to predict the category of the input image. The commonly explored classification problems in domain adaptation include depressive disorder identification, Alzheimer’s disease identification, cancer classification, and pneumonia diagnosis. Some commonly used classification datasets for domain adaptation are the Alzheimer’s Disease Neuroimaging Initiative (ADNI) \citep{jack2008alzheimer, weiner2017alzheimer}, CAMELYON17 \citep{litjens20181399}, TCGA dataset \citep{kandoth2013mutational}, and REST-meta-MDD Consortium \citep{yan2019reduced}. ADNI is a highly influential project in the research of Alzheimer’s disease (AD). It consists of four datasets: ADNI-1, ADNI-2, ADNI-GO, and ADNI-3. MRI, PET, and fMRI are the most popular modalities used in domain adaptation research within ADNI. CAMELYON17 and TCGA dataset are used for the classification of cancer tumor cells. CAMELYON17 is collected from five medical centers in the Netherlands and contains 1,399 annotated whole-slide images (including the CAMELYON16 dataset) of lymph nodes, both with and without metastases. The REST-meta-MDD Consortium dataset consists of R-fMRI indices from patients with Major Depressive Disorder (MDD) and matched normal controls (NCs). It includes 1300 MDDs and 1128 NCs.

\textbf{Image segmentation:}
Segmentation is a fundamental and necessary task in medical image analysis. It involves making pixel-by-pixel predictions to represent the morphology of biomedical structures such as cells \citep{xing2020bidirectional}, glands, and organs. Deep learning-based segmentation models can be classified into two types based on whether they distinguish each instance object: semantic segmentation and instance segmentation. Semantic segmentation aims to predict the category of each pixel to obtain object masks, which can be seen as a pixel-by-pixel classification task. In instance segmentation, models not only categorize pixels but also assign them a unique instance ID. The most commonly used dataset for cross-modality segmentation is the MM-WHS challenge dataset \citep{zhuang2016multi, zhuang2019evaluation}, which provides 120 multi-modality cardiac images. The BraTS dataset \citep{menze2014multimodal, bakas2018identifying}, MICCAI WMH Challenge dataset \citep{kuijf2019standardized}, Spinal cord gray matter segmentation (SCGM) dataset \citep{prados2017spinal}, Prostate MR Image Segmentation (PROMISE12) challenge dataset \citep{litjens2014evaluation}, and Calgary-Campinas-359 (CC-359) dataset \citep{souza2018open} are commonly used for segmentation tasks, where the domain shift is primarily caused by changes in institutions and scanners. The BraTS dataset is used for brain tumor segmentation and contains a total of 542 MR scans. The MICCAI WMH Challenge dataset has a total of 60 training and 110 test images. The SCGM challenge dataset is composed of 80 healthy subjects, split into 40 training and 40 test subjects, with 20 subjects acquired at each of the 4 different sites. The PROMISE12 dataset contains 100 prostate T2-weighted MRI cases from 4 different centers, and it is further subdivided into training (50), test (30), and live challenge (20) cases. The CC-359 dataset consists T1 volumes acquired in 359 subjects on scanners from three different vendors, with approximately 60 subjects per vendor and at two magnetic field strengths (1.5 T and 3 T). The MICCAI2018 IVDM3Seg dataset\footnote{https://ivdm3seg.weebly.com/} contains 3D multi-modality MRI data sets, with each set consisting of four aligned high-resolution 3D volumes. In total, there are 96 high-resolution 3D MRI volume data present in this dataset. The REFUGE challenge dataset \citep{orlando2020refuge} and the RIGA+ dataset \citep{hu2022domain} are used for the segmentation of the Optic Disc (OD) and Cup (OC). The REFUGE challenge dataset contains 1200 fundus images with ground truth segmentations and clinical glaucoma labels. The RIGA+ dataset is a combination of two different data sources and contains 290 unlabeled samples for the source domain and 1171 samples for the target domain, of which 454 samples are labeled.

\textbf{Others:} 
After image classification and segmentation tasks, image detection is a widely used task in medical image analysis. In contrast to classification, the detection task involves finding objects within entire images and determining the categories to which they belong. Therefore, it can be seen as a combination of two different tasks: regressing the location of the object and classifying the object's category. The commonly explored detection problems in domain adaptation include cell/nucleus detection \citep{xing2020bidirectional}, detection of breast cancer metastases, lesion detection (such as detecting polyps in colonoscopy images and detecting masses in mammography images) \citep{chen2022graphskt}, and colorectal cancer tissue detection \citep{abbet2022self}. Researchers often perform domain adaptation for image detection task by utilizing diverse datasets from different sources. The domain shift in these datasets is primarily caused by differences in staining procedures and imaging protocols/modalities. Apart from image detection, other tasks that have been addressed using domain adaptation include medical image synthesis \citep{hu2022domain}, depth estimation \citep{karaoglu2021adversarial, yu2021cross}, image restoration \citep{li2022annotation}, among others.

\begin{figure}[t]
\begin{center}
\includegraphics[scale=0.6]{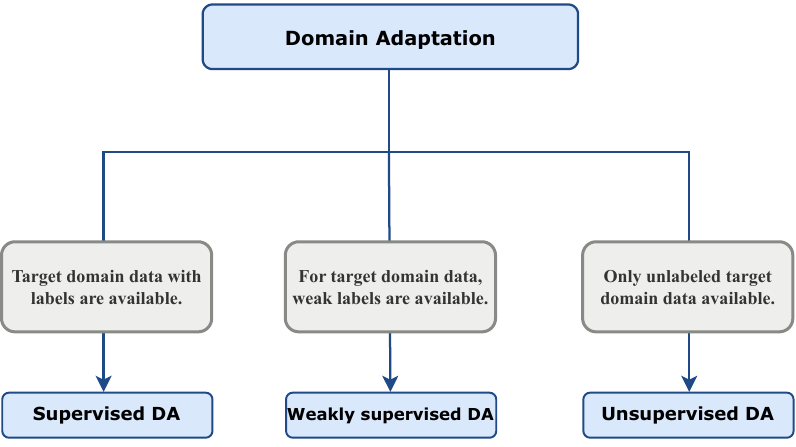}
\end{center}
\caption{Categorization of domain adaptation methods based on label availability.} \label{fig_2}
\end{figure}

\begin{table*}[t]
    \centering
    \scriptsize
    \begin{center}
  %  \tabstyle{5pt}
    \caption{Overview of deep supervised domain adaptation methods.}
    \label{table:supervised}
    \resizebox{\textwidth}{!}{
    \begin{tabular}{p{1cm}p{2.5cm}p{3cm}p{4.5cm}p{4cm}p{4.5cm}}\toprule
   Reference & Domain-alignment method
    & Task  & Dataset & Characterization of domain shift & Source domain $\rightarrow$ Target domain\\
    \midrule
    \citep{samala2018cross} & Two-stage Fine-tuning & Classifying masses in digital breast tomosynthesis & ImageNet \citep{deng2009imagenet} and  mass lesions collected from two imaging modalities: digitized-screen film mammography (SFM) and full-field digital mammography (DM) & Cross - domain dataset & ImageNet $\rightarrow$ SFM, SFM $\rightarrow$ DM\\
\midrule
\citep{gu2019progressive} & Fine-tuning  + Adversarial learning & Skin disease classification & ImageNet(I), MoleMap \url{(http://molemap.co.nz .)} and HAM(H) \url{(https://isic-archive.com/ .)} & Cross - domain dataset & I $\rightarrow$ H,
I $\rightarrow$ H+MA* (augmented version of molemap modality images),
I $\rightarrow$ H+MD* (augmented version of molemap dermoscopy) \\
\midrule
   \citep{ghafoorian2017transfer} & Fine-tuning & White matter hyper-intensity Segmentation & Radboud University Nijmegen Diffusion tensor and MR imaging Cohort (RUN DMC) \citep{van2011causes} MRI FLAIR images & Baseline scans and follow up scans were acquired with different voxel size and inter-slice gap & Baseline scans $\rightarrow$ Follow up scans \\
 \midrule

\citep{vesal2020automated} & Fine-tuning & Cardiac(Left ventricular cavity (LV) and Left ventricular myocardium (MYO) and right ventricle blood cavity (RV)) segmentation & MS-CMRSeg32019 \url{(http://www.sdspeople.fudan.edu.cn/zhuangxiahai/0/mscmrseg19/)} dataset (T2,  balanced-Steady State Free Precession (bSSFP) and Late Gadolinium Enhanced-MR (LGE) cardiac MR images & Large degree of variance in contrast and brightness &
 T2-weighted + bSSFP $\rightarrow$ LGE-MR subjects\\
 \midrule
 \citep{zakazov2021anatomy} & Fine-tuning & Skull stripping segmentation & Calgary-Campinas-359 (CC-359) \citep{souza2018open} & Acquired on scanners from three vendors (Siemens, Philips and General Electric)
& Siemens $\rightarrow$ Philips and General Electric \\
\midrule
\citep{han2018deep} & Fine-tuning & Projection-Reconstruction & CT data \url{(http://www.aapm.org/GrandChallenge/Low)}, Human Connectome Project (HCP) MR data \url{(https://db.humanconnectome.org)} and vivo radial MR datasets & Cross-modality and different target organs & CT $\rightarrow$ MRI\\
\midrule
\citep{feng2021deep} & DA based on underlying semantics of the training samples (shared weights) &
Pneumonia Diagnosis & ChestX-ray14 \citep{wang2017chestx} and Tan Tock Seng Hospital (TTSH) dataset for chest X-ray & Acquired from different institute & ChestX-ray14 $\rightarrow$ TTSH \\
\midrule
\citep{bermudez2018domain} & Two-stream architecture (MMD and correlation alignment) & 1.Synapses segmentation 2.Mitochondria segmentation & Private Transmission Electron Microscopy volumes of mouse brain & Different organs and stack size &1.Mouse cerebellum $\rightarrow$ Mouse somatosensory cortex \newline 2.Mouse striatum $\rightarrow$ Mouse hippocampus \\\midrule
\citep{laiz2019using} & Triplet loss & Classification of endoscopic images & Two datasets by using two different versions of the Wireless Capsule Endoscopy (WCE) capsules & Different camera (resolution quality) & Old WCE device $\rightarrow$ New WCE device \\
\midrule
\citep{wolleb2022learn} & Center point and latent loss & Classification task between multiple sclerosis (MS) patients and healthy controls & Private MS dataset and ADNI , Young Adult Human Connectome Project (HCP) \citep{Gerhard2013} and Human Connectome Project - Aging (HCPA) dataset for healthy controls
& Different scanner & ADNI + HCP + HCPA +Private MS dataset (Study 2, 3 and 4) $\rightarrow$ Private MS dataset (Study 1 and 5)\\
\midrule
\citep{sanchez2022cx} & PCA + CycleGan & Pneumonia diagnosis & Chest X-ray Images (Pneumonia) \url{(https://www.kaggle.com/ paultimothymooney/chest-xray-pneumonia)}
Dataset and private chest X-ray images  & Different medical center & Chest X-ray Images $\rightarrow$ Private chest X-ray images \\ 
\bottomrule

 \end{tabular}
 }
   \end{center}
\end{table*}

%%%%%%%%%%%%%%%%%%%%%%%%%%%%%%%%%%%%%%%%%%%%%%%%%%%%%%%%%%%%%%%%%%%%%%%%%%%%%%%%%%
\subsection{Overview of learning schemes}
\label{subsec_2.5}

In this section, we provide a formal introduction to various learning schemes in the context of deep learning applied to domain adaptation. Based on the availability of labels, existing domain adaptation methods can be categorized into supervised domain adaptation, weakly-supervised domain adaptation, and unsupervised domain adaptation, as shown in Fig.~\ref{fig_2}.

\subsubsection{Supervised domain adaptation (SDA) \label{subsubsec_2.5.1}}

In supervised domain adaptation (SDA), a small number of labeled data from the target domain are available for training the model. Typically, deep supervised domain adaptation techniques can be categorized into fine-tuning and discrepancy minimization methods (see Table~\ref{table:supervised}). Fine-tuning involves taking a pre-trained model, trained on a large dataset from the source domain, and adjusting its parameters or weights to better fit the target domain data. The hierarchical feature learning of Convolutional Neural Network (CNN) enables fine-tuning to be effective. The initial layers capture basic, universal visual building blocks such as edges, corners, and simple blob-like shapes, while the deeper layers learn more complex and abstract features specific to the task. The second approach, discrepancy minimization, utilizes two networks: one for the source domain and one for the target domain. Various loss functions or discrepancy measurements are employed to align the two domains and reduce their differences.

\textbf{Fine-tuning:} 
The most straightforward approach for supervised domain adaptation (SDA) is to utilize a pre-trained model trained on a raw image dataset and fine-tune it on the medical dataset. Several CNN-based methods have been proposed following this approach. For breast cancer diagnosis, Samala et al. \citep{samala2018cross} suggest using an AlexNet-like network pre-trained on the ImageNet natural image dataset and then fine-tuning it using regions-of-interest (ROI) from 2,454 mass lesions. In the case of skin cancer classification, Gu et al. \citep{gu2019progressive} first pre-train a CNN model on ImageNet, and then propose a two-step progressive transfer learning approach by successively fine-tuning the network on two skin disease datasets. Additionally, they employ adversarial learning as a domain adaptation technique to improve classification performance. Recently, Shamshiri et al. \citep{shamshiri2023compatible} propose a compatible-domain transfer learning approach to classify breast cancer cytological images into benign and malignant categories. Their approach involves pre-training the model using histopathological biopsy data, which exhibits patterns and structures that are somewhat comparable to the target cytological images. Tasks involving segmentation have also made use of fine-tuning. For example, Ghafoorian et al. \citep{ghafoorian2017transfer} investigate the effect of fine-tuning strategy on brain lesion segmentation using CNN models pre-trained on brain MRI scans. Their research shows that the transferability of models can be enhanced by fine-tuning them with a minimal amount of target training data. Based on this, several strategies have been devised to appropriately leverage CNNs pre-trained on a sizable dataset to solve medical imaging tasks. Vesal et al. \citep{vesal2020automated} perform supervised domain adaptation for multi-sequence cardiac MRI segmentation using fine-tuning. Regarding the fine-tuning process, there is an ongoing debate on which layers of the learned CNN should be fine-tuned. Some argue that fine-tuning early layers is more suitable for images with low-level domain shift, while others suggest fine-tuning deeper layers is more effective for images with high-level domain shift. To address this issue, Zakazov et al. \citep{zakazov2021anatomy} propose a CNN architecture that automatically selects the optimal layers for fine-tuning. In another application, Han et al. \citep{han2018deep} utilize fine-tuning to restore high-resolution MR images.

\textbf{Discrepancy minimization:} 
Another line of research in supervised domain adaptation (SDA) focuses on discrepancy minimization, which involves using two networks, one for the source domain and one for the target domain, along with various loss functions or discrepancy measurements to align the two domains \citep{feng2021deep, bermudez2018domain, shi2022improving}. Deep supervised DA (DSDA) \citep{feng2021deep} aims to transfer knowledge from the source domain's multi-label classification task to the target domain's binary pneumonia classification task. Based on the underlying semantics of the training samples, DSDA aligns the distributions of the source domain and the target domain. The feature extraction layers shared between these two sub-networks undergo end-to-end training. A two-stream U-Net architecture is designed for segmenting images from electron microscopy \citep{bermudez2018domain}. While one stream takes samples from the source domain, the other stream uses target data. Correlation alignment and Maximum Mean Discrepancy (MMD) are employed as domain regularization techniques for domain adaptation. In the context of multi-subject surface electromyography (sEMG)-based pattern recognition, Shi et al. \citep{shi2022improving} propose a CNN-based multi-task dual-stream SDA approach. This method utilizes two sub-networks based on CNN to improve the robustness and adaptability of the system. They adjust the weights of the CNN by combining gesture classification loss and domain variance loss. The domain variance loss minimizes the distribution divergence by aligning the feature distributions of the source and target domains.

To classify endoscopic images, Laiz et al. \citep{laiz2019using} employ triplet loss, where each triplet consists of a negative sample $C$ from the source domain, an anchor sample $A$ from the source domain, and a positive sample $B$ with the same label from the target domain. Their model aims to minimize the domain shift while ensuring discrimination among various diseases by reducing the triplet loss. In the context of MR images, Wolleb et al. \citep{wolleb2022learn} introduce specific additional constraints on the latent space to disregard scanner-related features. They also propose two novel loss terms that can be applied to any classification network. For pneumonia diagnosis, Sanchez et al. \citep{sanchez2022cx} propose a three-step approach. Firstly, they utilize Principal Component Analysis (PCA) subspaces to select the most representative images from the source domain. Secondly, they employ image-to-image translation based on a cycle Generative Adversarial Network (cycleGAN) to adapt the selected source domain samples to the target distribution. Finally, the modified source dataset images and the target training dataset are fed into a CNN for classifying the target test dataset.

\begin{table*}[t]
    \centering
    \scriptsize
    \begin{center}
  %  \tabstyle{5pt}
    \caption{Overview of deep weakly-supervised domain adaptation methods.}
    \label{table:weakly-supervised}
   \resizebox{\textwidth}{!}{
    \begin{tabular}{p{1cm}p{2.5cm}p{3cm}p{4.5cm}p{4cm}p{4.5cm}}
    \toprule
   Reference & \multicolumn{1}{l}{Domain-alignment method} & \multicolumn{1}{l}{Task} & \multicolumn{1}{l}{Dataset} & \multicolumn{1}{l}{Characterization of domain shift} & \multicolumn{1}{l}{Source domain $\rightarrow$ Target domain}\\ \midrule 

\multicolumn{2}{l}{\textbf{Semi-Supervised Methods}} &&&& \\ \midrule
\citep{roels2019domain} & Encoder decoder architecture and a reconstruction decoder is used to align the features & Mitochondria segmentation in electron microscopy volumes & Source dataset consists of two annotated 165 × 1024 × 768 FIB-SEM acquisitions, HeLa cell and Drosophila dataset \citep{Gerhard2013} & Source and HeLa cell dataset: different acquisition parameters, source and Drosophila dataset: different modality & Source $\rightarrow$ HeLa cell dataset, Source $\rightarrow$ Drosophila dataset \\ \midrule

\citep{xu2021cross} & Representation learning & Assess COVID-19 disease using CT scans & Private CT images and MosMedData \citep{morozov2020mosmeddata} & Private CT images acquired from three clinical sites (Site-1, Site-2, and Site-3) and MosMedData (site 4) & Site-1 (training), Site-2 $\rightarrow$ Site-1 (test), Site-2 (training), Site-1 $\rightarrow$ Site-2 (test), Site-3 (training), Site-1 $\rightarrow$ Site-3 (test), Site-4 (training), Site-1 $\rightarrow$ Site-4 (test)\\ \midrule

\citep{madani2018semi} & Adversarial learning & Cardiac abnormality classification in chest X-rays &
NIH PLCO dataset \citep{oken2011screening} and NIH Chest X-Ray \citep{demner2016preparing} & Different medical center & NIH PLCO dataset $\rightarrow$ NIH Chest X-Ray \\ \midrule

\citep{liu2022act} & Pseudo-labeling & Brain tumor MRI segmentation & BraTS2018 database
& Different scanners & T2-weighted $\rightarrow$ T1-weighted/T1ce/FLAIR \\ \midrule

\citep{fotedar2020extreme} & Extreme consistency &  Fundus image segmentation & Fundus image datasets: HRF and STARE & Different scanners and pathologies & HRF $\rightarrow$ STARE \\ \midrule

\citep{li2020dual} & Pseudo-labeling (dual-teacher) & Cardiac (RVC, LAC, LVC, RAC, MYO, AA and PA) segmentation
 & MM-WHS challenge dataset (2017) & Cross-modality dataset & MRI $\rightarrow$ CT \\ \midrule

 \citep{li2020dual} & Pseudo-labeling (dual-teacher++) & Cardiac (RVC, LAC, LVC, RAC, MYO, AA and PA) segmentation & 
MM-WHS challenge dataset (2017) & Cross-modality dataset & MRI (CT) $\rightarrow$ CT (MRI) \\ \midrule

\citep{gu2022contrastive} & Pseudo-labeling (teacher-student network) & 1.segmentation of coronary  arteries in X-ray images \newline 2.Segmentation: LV and MYO from MRI images & DRIVE \url{(https://drive.grand-challenge.org/)}, REFUGE, MS-CMRSeg and X-ray dataset for coronary arteries & Different organs and cross-modality but same anatomical structures of the organ & 1.DRIVE (fundus images) $\rightarrow$ private dataset (coronary arteries) \newline 2. REFUGE $\rightarrow$ MS-CMRSeg \\ \bottomrule

\multicolumn{2}{l}{\textbf{Incomplete Supervision Methods}} &&&& \\ \midrule

\citep{dong2020weakly} & Adversarial learning &  Endoscopic Lesions Segmentation & Medical Endoscopic Dataset \citep{dong2019semantic} & Different organs & Gastroscope samples $\rightarrow$ Enteroscopy samples \\ \midrule

\citep{yang2021minimizing} & Adversarial learning and Pseudo labels  & Nucleus instance segmentation and classification & Colorectal nuclear
segmentation and phenotype (CoNSep - single cancer type) \citep{graham2019hover} and PanNuke \citep{gamper2019pannuke} (19 cancer types) & Cross domain images & CoNSep $\rightarrow$ PanNuke \\ \midrule

\citep{bateson2021constrained} & Domain-invariant prior knowledge 
& 1.Lower spine segmentation \newline 2.Cardiac(AA, MYO, RV and LV) segmentation & 1.MICCAI 2018 IVDM3Seg challenge  \newline 2.MMWHS Challenge dataset & 1.Different Modalities (in-phase, opposed-phase, fat and water images) \newline 2.Cross-modality dataset & 1.Water modality $\rightarrow$ In-Phase modality \newline 2. MRI $\rightarrow$ CT \\ \midrule

\citep{wang2021deep} & Adversarial learning & Breast cancer screening from mammograms & Curated Breast Imaging Subset of the Digital Database for Screening Mammography (CBIS-DDSM) and Dataset of Breast Screening Mammography from the West China Hospital (DBSM-WCH) & Different institutes & CBIS-DDSM $\rightarrow$ DBSM-WCH \\ \midrule

\citep{cao2022collaborative} & Multi-instance learning + CycleGAN & Classification: diabetic retinopathy (DR) grading & Messidor dataset \citep{decenciere2014feedback} and large-scale Eyepacs dataset (IDRiD) & Different sources and annotations & IDRiD $\rightarrow$ Messidor dataset  \\ 

\bottomrule
\end{tabular}
}
\end{center}
\end{table*}

%%%%%%%%%%%%%%%%%%%%%%%%%%%%%%%%%%%%%%%%%%%%%%%%%%%%%%%%%%%%%%%%%%%%%
\subsubsection{Weakly-supervised domain adaptation (WSDA) \label{subsubsec_2.5.2}} 

Weakly supervised learning refers to a broad range of studies that aim to construct predictive models using learning under weak supervision. Weak supervision can be classified into three common types \citep{zhou2018brief}. The first type is incomplete supervision, where labels are provided for only a subset of the training data. The second type is inexact supervision, where the training data is annotated with coarse-grained labels that provide limited detail. The third type is inaccurate supervision, where the provided labels may not always accurately represent the ground truth and can contain errors or inaccuracies. Weakly-supervised domain adaptation techniques strive to bridge the domain gap between the source and target domains while effectively utilizing the limited or imperfect supervision available in the target domain to train the model (see Table~\ref{table:weakly-supervised}).

\textbf{Incomplete supervision or semi-supervised domain adaptation:} 
For electron microscopy segmentation, Roels et al. \citep{roels2019domain} propose a novel approach that incorporates a reconstruction decoder into the traditional encoder-decoder segmentation framework. This addition aims to align the encoder features of both the source and target domains. They initially train the network using unsupervised training. After unsupervised training, the reconstruction decoder is removed, and the entire network is fine-tuned using labeled target samples to ensure adaptation to the target domain. Xu et al. \citep{xu2021cross} address the class imbalance problem and mitigate domain discrepancy by employing representation learning with three loss functions. They utilize the prototype triplet loss to transfer distinguishing feature knowledge from the source domain to the target domain. Furthermore, they employ the conditional maximum mean discrepancy (CMMD) loss and a multi-view reconstruction loss on the representation to enhance the discriminative power and ensure the completeness of the latent space. For the assessment of COVID-19 disease using CT scans collected from various centers, a semi-supervised GAN-based architecture \citep{madani2018semi} is utilized by adapting the generator to work with both labeled and unlabeled data. As the model converges, the discriminator becomes proficient at distinguishing between generated images representing diseases and real images depicting either diseases or normal conditions. This architecture addresses both the problems of labeled data scarcity and data domain overfitting. The experiments are conducted for cardiac abnormality classification in chest X-rays.

In recent advancements in deep semi-supervised methods, consistency learning has emerged as a prominent approach. This methodology typically involves two key roles: a teacher model and a student model, forming a Teacher-Student framework. In this framework, the teacher model is often implemented as an exponential moving average of the student model, and the teacher assists the student in approximating its performance in the presence of perturbations. A few works follow the student-teacher paradigm to address semi-supervised domain adaptation \citep{fotedar2020extreme,li2020dual,li2020dual,gu2022contrastive}. Fotedar et al. \citep{fotedar2020extreme} introduce an approach that leverages a teacher-student paradigm to achieve extreme consistency. The student network is presented with an extreme variation of an image, and its prediction should align with the teacher's prediction for the original image. They evaluate the method on skin lesion and two retinal image datasets. Dual-teacher \citep{li2020dual} is a student-teacher model in which the student model learns from limited labeled target data and is supervised by two teachers. The inter-domain teacher uses labeled source data to guide the student model, whereas the intra-domain teacher uses unlabeled target data to provide guidance, yielding impressive results for cardiac segmentation. Dual-teacher++ \citep{li2020dual+} is an extension of the earlier approach with the dual-domain reliability control strategy, which reduces uncertain transfer and promotes reliable intra- and inter-domain knowledge integration. Gu et al. \citep{gu2022contrastive} propose a domain-specific batch normalization segmentation network with shared convolutional kernels that learns from labeled source and target data to provide supervision and resolve the cross-anatomy gap. Furthermore, a self-ensembling mean teacher module is utilized with unlabeled target data to improve prediction. A consistency loss is applied to the student and teacher predictions over different perturbations of the same image. The performance is evaluated on datasets with the same anatomical structure but potentially different scales. Asymmetric co-training (Act) for cross-modality brain tumor MRI segmentation \citep{liu2022act} utilizes two segmentors: the first segmentor performs a traditional UDA task using labeled data from the source domain and unlabeled data from the target domain, while the second segmentor performs semi-supervised learning (SSL) task using both labeled and unlabeled data from the target domain. The knowledge acquired from these two modules is then adaptively merged with Act by iteratively teaching one another using the confidence-aware pseudo-label. Moreover, an exponential MixUp decay technique is employed to effectively control pseudo label noise and facilitate smooth propagation.

\textbf{Inexact supervision:} 
With only image-level annotations, Dong et al. \citep{dong2020weakly} aim to develop a pixel-level endoscopic segmentation model. They utilize adversarial learning to mitigate the domain gap between source gastroscope samples and target enteroscopy data. Furthermore, to align category-wise feature centroids, they create a self-supervised pseudo-pixel label generator with class balance and super-pixel prior. Moreover, a quantified transferability approach built on a Wasserstein adversarial network is designed to investigate transferable contextual dependencies while ignoring irrelevant semantic representation. To minimize the labeling cost, Yang et al. \citep{yang2021minimizing} propose a novel application that treats each cancer type as a distinct domain and employs domain adaptation techniques to enhance the performance of segmentation and classification across different cancer types. They propose an integrated approach that combines unsupervised domain adaptation and weakly supervised domain adaptation, suitable for different levels of annotations, including point-level, image-level, and no annotations. Cyclic adaptation with pseudo labels and an adversarial discriminator are utilized for unsupervised domain alignment. Moreover, image-level or point-level annotations are employed to supervise nucleus classification and refine the pseudo labels. In their work, Bateson et al. \citep{bateson2021constrained} introduce a constrained framework for domain adaptation that focuses on enforcing image-level statistics, particularly anatomical information, in the target domain. These statistics can be learned from the source domain or known a priori. Furthermore, weak annotations of the target data, such as image-level tags, can be incorporated using inequality restrictions. Remarkably, even when using only image-level annotations in the target domain, their method achieves performance comparable to fully supervised approaches for cross-modality segmentation tasks. To align domains annotated at different levels, Wang et al. \citep{wang2021deep} propose a two-stage method. The first stage involves domain adaptation, where the source domain with image-level labels and the target domain with no labels are aligned using adversarial learning. In the second stage, the target data is labeled at the case-level, meaning each sample of the target data consists of two images that share a single label. To leverage the case-level precision labels and fine-tune the entire network, the authors incorporate a feature fusion component that reshapes the features of each case. The effectiveness of this method is demonstrated in the context of breast cancer screening. Cao et al. \citep{cao2022collaborative} present a unified framework for weakly-supervised domain adaptation. This framework comprises three main components: domain adaptation, instance progressive discriminator, and multi-instance learning with attention. The authors evaluate the effectiveness of their framework on the Messidor dataset \citep{decenciere2014feedback} and the large-scale Eyepacs dataset for the task of diabetic retinopathy (DR) grading.

\textbf{Inaccurate supervision:} Inaccurate supervision refers to the scenario where the provided supervision information is not entirely accurate and may contain errors \citep{zhou2018brief}. While there have been research efforts addressing both domain adaptation and supervision inaccuracy together \citep{yu2020label}, it is worth noting that in the domain of medical imaging, these two challenges have not been effectively addressed in a unified manner.

\subsubsection{Unsupervised domain adaptation (UDA) \label{subsubsec_2.5.3}}

Deep unsupervised domain adaptation refers to a technique that aims to adapt a model trained on a source domain to perform effectively on a target domain without needing labeled target domain data \citep{guan2021domain}. In the field of medical imaging, considerable research has been conducted on UDA. Various methods have been proposed to address UDA challenges in medical imaging \citep{ganin2015unsupervised, zhu2017unpaired, yang2019unsupervised}. In Section~\ref{sec_3}, we will provide an overview of different approaches and techniques employed for UDA in medical imaging.

\subsection{Related topics \label{subsec_2.4}}
In this section, we discuss the connections and differences between DA and its related topics.

\textbf{Transfer learning (TL):}  
Domains are described as the union of an input space $X$, an output space $Y$, and a corresponding probability distribution $P$. Inputs, also known as feature vectors or points in feature space, are subsets of the D-dimensional real space $\mathbb{R}^D$. The outputs are classes, where $Y$ can be binary or multi-class. When two domains are compared, they are said to be different if they differ in at least one of their $X$, $Y$, or $P$ constituent parts. The definition of TL refers to the broad situation where the domains are free to differ in sample space, label space, distribution, or all of these \citep{kouw2018introduction}. Examples of differences across feature spaces include how image caption generators from vision model generalize from the ``image domain" to the ``text domain" \citep{gopalan2015domain, karpathy2015deep}. DA is the specific situation where only the probability distributions change but the sample and label spaces remain the same. Fine-tuning is a well-known TL example in modern deep learning: deep neural networks are first pre-trained on massive datasets, such as ImageNet \citep{deng2009imagenet} for image models, and are then fine-tuned on downstream tasks \citep{girshick2014rich}.

\textbf{Domain generalization (DG):} 
In DG \citep{zhou2022domain, li2018learning, 9961940, hua2023dcam}, the objective is to learn a model using data coming from one or multiple related but separate source domains in a way that the model generalizes well to any out-of-distribution (unseen) target domain. The main distinction between DA and DG depends on whether the target data is utilized or not. In DG, we make the assumption that we don't have access to the target data, putting a greater emphasis on model generalization. DA, on the other hand, implies that sparsely labeled \citep{shamshiri2023compatible} or unlabeled target data \citep{huang2022domain} are available for model adaptation, giving it access to the marginal distribution of target data.

\begin{figure}[t]
\begin{center}
\includegraphics[scale=0.9]{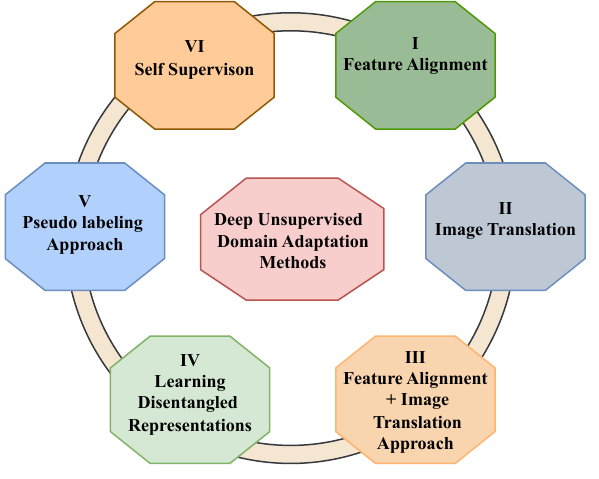}
\end{center}
\caption{Categories of deep unsupervised domain adaptation methods.} \label{fig_3}
\end{figure}

\section{Deep unsupervised domain adaptation methods \label{sec_3}}

This section presents the concepts, formulations, and general procedures of deep UDA methods. We divide this section into: feature alignment (Section~\ref{subsec_3.1}), image translation-based methods (Section~\ref{subsec_3.2}), image translation + feature alignment methods (Section~\ref{subsec_3.3}), pseudo-labeling based methods (Section~\ref{subsec_3.4}), disentangled representations methods (Section~\ref{subsec_3.5}), and self-supervised methods (Section~\ref{subsec_3.6}) as shown in Fig.~\ref{fig_3}. We also discuss the respective datasets used in the UDA methods to assess the divergence between the different domains in Table~\ref{table:Unsupervised}.

\subsection{Feature alignment \label{subsec_3.1}}

\begin{figure}[t]
\begin{center}
\includegraphics[scale=0.42]{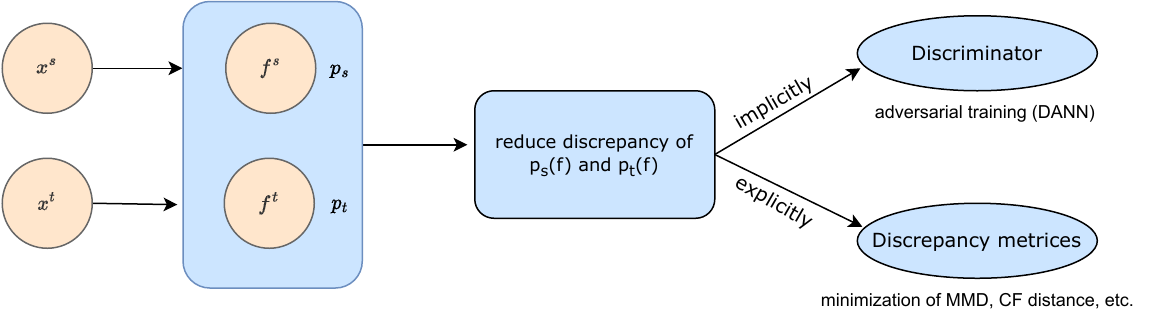}
\end{center}

\caption{Framework for domain adaptation research in the feature space. The domains are mapped into a common latent feature variable $f$. The domain discrepancy is then reduced using either adversarial training or explicit minimization of discrepancy measures.}

\label{fig_4}
\end{figure}

The core concept of feature alignment in unsupervised domain adaptation is to reduce the disparity between the source and target domains by learning domain-invariant representations. Many UDA approaches aim to map images from both domains into a shared latent space to minimize discrepancies. This can be achieved explicitly by minimizing a discrepancy metric that quantifies the difference between the domains, implicitly through adversarial learning techniques (see Fig.~\ref{fig_4}) and by graph based methods. The goal is to align the feature distributions of the source and target domains to ensure that the learned representations are transferable and effective across domains.

\subsubsection{Explicit discrepancy minimization}
Explicit discrepancy minimization methods typically define a discrepancy metric or loss function that quantifies the dissimilarity between the source and target distributions. This metric is minimized during training to encourage domain-invariant feature learning. Various discrepancy metrics, such as Maximum Mean Discrepancy (MMD), Kullback-Leibler (KL) divergence, and Contrastive Loss (CL), can be used. 

\textbf{Classification:} 
The Maximum Mean Discrepancy (MMD) calculates the distance between two probability distributions by initially mapping inputs to a reproducing kernel Hilbert space (RKHS) and then computing the discrepancy based on their means. Some studies have built UDA methods based on explicitly minimizing the MMD metric. For instance, Yu et al. \citep{yu2022domain} utilize two parallel feature encoders for the source and target domains. They incorporate an attention mechanism to capture specific brain regions and utilize MMD to learn domain-invariant features for cross-modality alignment. Similarly, Fang et al. \citep{fang2023unsupervised} align the extracted features from the attention-guided spatio-temporal graph convolution module using MMD for major depressive disorder classification. To predict COVID-19 malignant progression and improve the model's generalization, Fang et al. \citep{fang2021deep} perform UDA in a multi-center study. They first pre-train the model on source data and then adapt it using a metric-based approach, where the prototype representations learned from the source center are passed to the target center.

\textbf{Segmentation:} 
Minimizing contrastive loss is one way of achieving domain alignment. By training the network to map input samples from different domains into a shared feature space, the contrastive loss aims to maximize the similarity between instances from the same domain and minimize the similarity between instances from other domains. One key consideration in contrastive loss is how to generate image pairs. The motivation behind using contrastive loss and the design of loss functions stem from the observation that object shape is a consistent feature across both domains \citep{sahu2020endo, gomariz2022unsupervised}. In their work, Sahu et al. \citep{sahu2020endo} employ two perturbation strategies: one involves changing pixel intensities, while the other is based on pixel corruption. They jointly train the network using a supervised loss for simulated labeled data and a consistency loss for real unlabeled data. Their framework is validated for the instrument segmentation task. For Optical Coherence Tomography (OCT) segmentation, Gomariz et al. \citep{gomariz2022unsupervised} propose a novel pair generation strategy that leverages the coherence of neighboring slices in a 3D volume. They jointly train the network using both supervised and contrastive loss. Another explicit metric for UDA, the Characteristic Function (CF) distance \citep{wu2020cf}, calculates the distance between the latent feature distributions in the frequency domain rather than the spatial domain. The authors conduct experiments on two medical image segmentation datasets.

Some studies encourage the distributions of the domains to approximate a specific parameterized probability distribution function, typically a normal distribution. This is achieved by utilizing an autoencoder to facilitate the approximation process \citep{wu2021unsupervised} \citep{lu2021learning}. Wu et al. \citep{wu2021unsupervised} perform a transformation of each domain into a latent feature variable. The two variables are then driven and approximated by a common and parameterized variational form via variational autoencoder (VAEs). Since this approximation can be estimated using either the source or target data, two estimations are obtained for the distribution of the common variational form. The authors leverage the distance between these two estimations as an effective regularization technique for domain adaptation. In addition to using VAE, Al Chanti and Mateus \citep{al2021olva} utilize the optimal transport theory (OT) loss to match and align the remaining discrepancies between the two domains in the latent space. They design a shared latent space that focuses on modeling shape rather than intensity variations, which contributes to the successful segmentation process.

\textbf{Others:} In their work, Hu et al. \citep{hu2022domain} utilize KL divergence as a measure for domain alignment. They introduce UDA for medical image synthesis in a 3D manner, employing a 2D VAE. To address the discrepancy between the source and target domains in image synthesis, they minimize the KL divergence between the posteriors of the generated outputs. This is necessary as the output spaces of the source and target domains differ. Lu et al. \citep{lu2021learning} propose the utilization of a biomechanically constrained autoencoder network to learn the latent representation of noisy displacements. To ensure the acquisition of a meaningful representation, the authors impose limitations on the autoencoder by incorporating prior information that well-regularized displacement patches should adhere to biomechanical constraints. The experiment is conducted for cardiac strain analysis, where synthetic data is used as the source domain and in vivo data serves as the target domain.

\subsubsection{Implicit discrepancy minimization}

\begin{figure}[t]
\center
 \includegraphics[width= 1\linewidth]{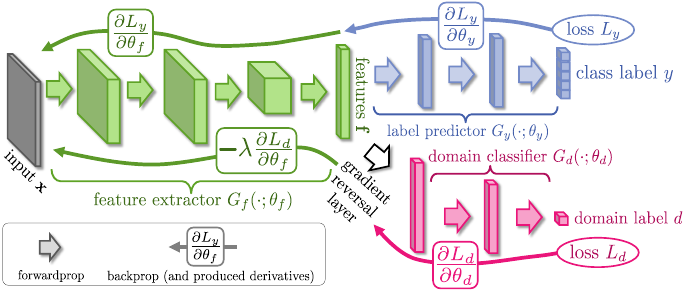}

 \caption{Illustration of the Domain Adversarial Neural Network (DANN) framework, a classic and efficient model for domain-invariant feature learning through adversarial training. Image courtesy of Ganin \citep{ganin2016domain}.}
 \label{fig_5}
\end{figure}

Implicit discrepancy minimization methods in UDA primarily rely on the principles of adversarial learning \citep{NIPS2014_5ca3e9b1}. This approach involves training two models: a generative model G, which captures the data distribution, and a discriminative model D, responsible for binary label prediction indicating whether a sample belongs to the training dataset or is generated by G. The training process employs a mini-max approach, where the label prediction loss is simultaneously optimized by minimizing G's loss and maximizing D's likelihood of correctly assigning labels. In UDA, this technique ensures that the network cannot differentiate between the source and target domains. To enforce comparability of feature distributions across domains, the domain-adversarial neural network (DANN) \citep{ganin2015unsupervised} incorporates a gradient reversal layer (GRL) into the Generative Adversarial Network (GAN) framework, as ahown in Fig.~\ref{fig_5}. The network consists of two classifiers and shared feature extraction layers. DANN utilizes GRL to maximize domain confusion loss while minimizing label prediction loss for source samples and domain confusion loss for all samples. DANN serves as a base model for various UDA methods based on adversarial learning.

\textbf{Classification:} 
To improve the classification of whole slide images, Ren et al. \citep{ren2018adversarial} employ a siamese architecture in the target domain and combine it with an adversarial loss for regularization purposes. Zhang et al. \citep{zhang2019whole} use adversarial learning to reduce the inter-domain discrepancy and entropy loss to enhance the similarity between classes (resolution, scale, color, etc.), and introduce focal loss to address the class imbalance problem in histopathology cancer images. The attention-guided deep domain adaptation (ADA) framework \citep{guan2021multi} consists of three key components: a feature encoding model for MRI feature extraction, an attention discovery module for identifying disease-related regions in brain MRIs, and a domain transfer module trained using adversarial learning to transfer knowledge between the source and target domains. They show the effectiveness of the method for the task of multi-site MRI harmonization. Recently, Feng et al. \citep{feng2023contrastive} perform binary and multi-class classification tasks to diagnose pneumonia. They employ a conditional domain adversarial network to reduce the domain gap and utilize a contrastive loss to address the issue of limited data in the target domain. Cai et al. \citep{cai2023prototype} propose a multi-scale feature extraction module for 3D MRI to handle MRI data with a domain shift problem and achieve automatic auxiliary diagnosis of alzheimer's disease. In their work, Zhang et al. \citep{zhang2023gionet} introduce a novel Global Information Optimized Network for multi-center COVID-19 diagnosis called GIONet. It incorporates a novel coronavirus disease GAN (COVID-GAN) and a 3D CNN with a graph-enhanced aggregation unit and a multi-scale self-attention fusion unit to enhance the global feature extraction capability.

\textbf{Segmentation:} 
The key idea in \citep{dou2018unsupervised} is that cross-modality domains (MRI/CT data) exhibit a significant gap in low-level characteristics (e.g., gray-scale values) rather than in high-level traits (e.g., anatomical structures). The authors implement this idea by updating the earlier layers, which learn the low-level features, while reusing the learned features in the higher layers of the CNN. The authors expand on the methodology suggested in \citep{dou2018unsupervised} by incorporating an additional discriminator to align segmentation masks from the source and target domains \citep{dou2019pnp}. Dou et al. \citep{dou2018unsupervised, dou2019pnp} conducted experiments on cross-modality cardiac images and achieved superior results. Attention mechanisms primarily emphasize disease-related regions in images while suppressing irrelevant details. Liu et al. \citep{liu2021automated} utilize the attention mechanism alongside adversarial learning to improve the accuracy of cardiac segmentation. In entropy-guided UDA, the source model is trained in a supervised manner, assigning low entropy values to source domain images and high entropy values to target domain images. To bridge the entropy gap between the two domains, a specific discriminator is employed \citep{zeng2020entropy, wang2019boundary}. Specifically, Zeng et al. \citep{zeng2020entropy} utilize a feature discriminator and an entropy-aligning discriminator. The entropy-aligning discriminator ensures that the target images possess low entropy values similar to the source domain images. Similarly, Wang et al. \citep{wang2019boundary} employ an entropy-aligning discriminator and another discriminator to align the boundary regions of target and source images, thereby enhancing the performance of OD and OC segmentation, especially in challenging ambiguous boundary regions.

Few studies have employed a combination of adversarial learning and self-training for medical image segmentation \citep{bian2020uncertainty, liu2021s, huang2022domain, li2022domain}. Specifically, Bian et al. \citep{bian2020uncertainty} utilize self-training and adversarial training schemes to align the features extracted from different domains by updating the shared uncertainty map estimation parameters. They conducted experiments on private cross-device OCT and public cross-modality cardiac datasets.
The Self-cleansing UDA (S-cuda) method \citep{liu2021s} addresses the domain shift problem and handles noisy labels in the source domain. Their approach employs self-training to generate precise pseudo-labels for the noisy source and unlabeled target domains.
To improve the generalization capability of the model for mitochondria segmentation, Huang et al. \citep{huang2022domain} focus on addressing the inter-section and intra-section gaps between the source and target domains. They propose a method that utilizes a CNN to learn an inter-section residual based on the segmentation results of adjacent sections from both domains. Furthermore, they use adversarial learning to align the inter-section residuals and leverage the learned inter-section residual to generate pseudo-labels, which are then employed to supervise the segmentation process.

In histopathology images, nuclei objects belong to different classes and exhibit distinct characteristics. To address this, Li et al. \citep{li2022domain} propose the category-aware feature alignment (CA) method. They first design a CA module with dynamic learnable trade-off weights. Then, to improve the model's performance on target data, they propose self-supervised training using pseudo-labels derived from nuclei-level prototype features. The experiments are conducted on the Lizard dataset \citep{graham2021lizard}, focusing on nuclei instance segmentation and classification. To handle the issue of large variance within categories and minor differences between categories, Liu et al. \citep{liu2022margin} propose a margin-preserving contrastive learning method. They introduce a penalty in the form of a deviation angle to preserve the margin for positive anchors and employ adversarial learning to align the entropy of both domains.

To address the limitations of domain adversarial training with a small number of target samples, Ouyang et al. \citep{ouyang2019data} propose a method that combines prior regularization on a shared feature space with domain adversarial training. The process involves utilizing a VAE-based feature prior matching approach, which is data-efficient. This combined approach aims to learn a shared latent space that is invariant to domain variations and can be effectively leveraged during the segmentation process. Feng et al. \citep{feng2022unsupervised} propose a novel paradigm for UDA that incorporates category-level regularization. Their approach involves aligning global distributions between domains through adversarial learning. Additionally, they address fine-grained level category distribution adaptation from three perspectives: the source domain, the target domain, and the inter-domain, utilizing three different category regularization methods.

Despite residing in low-dimensional space, segmentation outputs contain rich information, including context and scene layout. The segmentation of images from both the source and target domains should exhibit high spatial and local similarities \citep{tsai2018learning}. To address this, Tsai et al. \citep{tsai2018learning} employ an adversarial learning strategy to adapt the low-dimensional softmax segmentation prediction outputs. By assuming that the segmentation outputs are domain-invariant, they aim to ensure that the segmentation network produces target domain predictions that closely resemble the source domain predictions \citep{haq2020adversarial}. For joint OD and OC segmentation, Wang et al. \citep{wang2019patch} uses patch-based adversarial learning to align the output space. Further, they employ a lightweight and effective network along with a morphology-aware segmentation loss to generate precise predictions. Haq and Huang \citep{haq2020adversarial} also leverage adversarial learning to align the output space. They additionally utilize a decoder network to maximize the correlation between target predictions and target images. This strategy is also extended to semi-supervised settings. Collaborative Feature Ensembling Adaptation (Cfea) \citep{liu2019cfea} employs feature ensembling and involves three networks: one for the source domain and two for the target domain (teacher and student networks). Each network plays a unique role in facilitating the learning of domain-invariant representations. Cfea makes use of adversarial discriminative learning in intermediate representation and output spaces to enhance feature adaptation.

\textbf{Others:} 
Some approaches \citep{karaoglu2021adversarial, yu2021cross} utilize UDA for depth estimation. For bronchoscopic depth estimation, Karaoglu et al. \citep{karaoglu2021adversarial} propose a two-step framework that involves supervised training of a depth estimation network using labeled synthetic images, followed by the adoption of an unsupervised adversarial domain feature adaptation scheme to enhance performance on real images. For cross-domain depth estimation, Yu et al. \citep{yu2021cross} propose a method that leverages detailed 2D OCTA depth maps instead of 3D volumetric data. They first estimate the depth map of blood vessels and then introduce a depth adversarial adaptation module for enhanced unsupervised cross-domain training. By utilizing the computed depth map and 2D vascular data, they reconstruct the vessels in a 3D space.
User-guided domain adaptation (UGDA) \citep{raju2020user} is a valuable strategy for quickly annotating 3D medical volumes through user interaction. It uses UDA to integrate the mask predictions with user interactions. A novel Sparse-based domain Adaptation Super-Resolution network (SASR) \citep{hao2022sparse} is proposed for the reconstruction of accurate low-resolution (LR) OCTA images to high-resolution (HR) representations. It employs a generative-adversarial technique to guide the reconstruction of LR images, allowing the unification of synthetic and real LR images in the feature domain. Multilevel Structure-Preserved GAN (MSP-GAN) \citep{xia2022multilevel} is an intravascular ultrasound (IVUS) DA module designed for transferring IVUS domains while meticulously conserving intravascular features. On the adversarial learning baseline, the MSP-GAN incorporates the transformer, contrastive restraint, and self-ensembling methods to efficiently preserve structures at several global, local, and fine levels. The effectiveness of the MSP-GAN in maintaining structures and assuring downstream accurate segmentation and quantification is evaluated. Automated sleep staging aids in evaluating treatment effectiveness and sleep quality. Yoo et al. \citep{yoo2021transferring} propose UDA for automated sleep staging. They improve upon the classical domain discriminator by employing local discriminators (subject and stage) to preserve the inherent structure of sleep data and reduce local misalignments.

\subsubsection{Graph based methods}
The Graph Convolutional Network (GCN) \citep{kipf2016semi} is a powerful method for understanding and representing relationships in graph structures. In recent years, it has been applied to various computer vision tasks and has achieved remarkable results. More recently, researchers have extended the use of graph techniques to develop frameworks that align the distributions between two domains at the feature level.

\textbf{Classification:} 
To diagnose neurodevelopmental disorders, Wang et al. \citep{wang2020unsupervised} investigate the use of UDA on graph domains. Their approach consists of three main components: graph isomorphism encoders, progressive feature alignment (PFA), and unsupervised infomax regularizer (UIR) \citep{wang2020unsupervised}. The PFA module aims to gradually and reliably align graph representations of both domains, while the UIR component enhances the feature alignment by learning effective unsupervised graph embeddings.

\textbf{Segmentation:} 
Some studies introduce graph-based UDA for surgical instrument segmentation \citep{liu2021prototypical,liu2021graph}. Specifically, Liu et al. \citep{liu2021prototypical} design a module that generates multi-prototypes to accurately capture semantic information, which is then shared within each domain. They employ hierarchical graphs based on these multi-prototypes to minimize the domain gap and utilize dynamic reasoning to transfer correlated information between the two domains \citep{liu2021prototypical}. In their work, Liu et al. \citep{liu2021graph} introduces three main modules to reduce the domain gap. The first module aggregates features into domain-specific prototypes and facilitates information sharing among these prototypes. The second module guides the evolution of feature maps towards a domain-common direction. Lastly, the third module evaluates category-level alignment from a graph perspective and assigns different adversarial weights to each pixel. This approach further enhances the alignment of feature distributions and improves overall performance in reducing the domain gap. 

\textbf{Others:} In their framework, Chen et al. \citep{chen2022graphskt} propose a novel approach for the domain adaptive lesion detection task. They leverage graph structures to model both intra-domain and inter-domain interactions, allowing for a comprehensive analysis of the relationships within each domain and across different domains. By incorporating graph-based representations, the model can effectively capture the contextual information and dependencies between lesions and their surrounding regions.

%%%%%%%%%%%%%%%%%%%%%%%%%%%%%%%%%%%%%%%%%%%%%%%%%%%%%%%%%%%%%%%%%%%%%%%%%%%%%%%

\subsection{Image translation based methods \label{subsec_3.2}}

\begin{figure}[t]
\begin{center}
\includegraphics[scale=0.35]{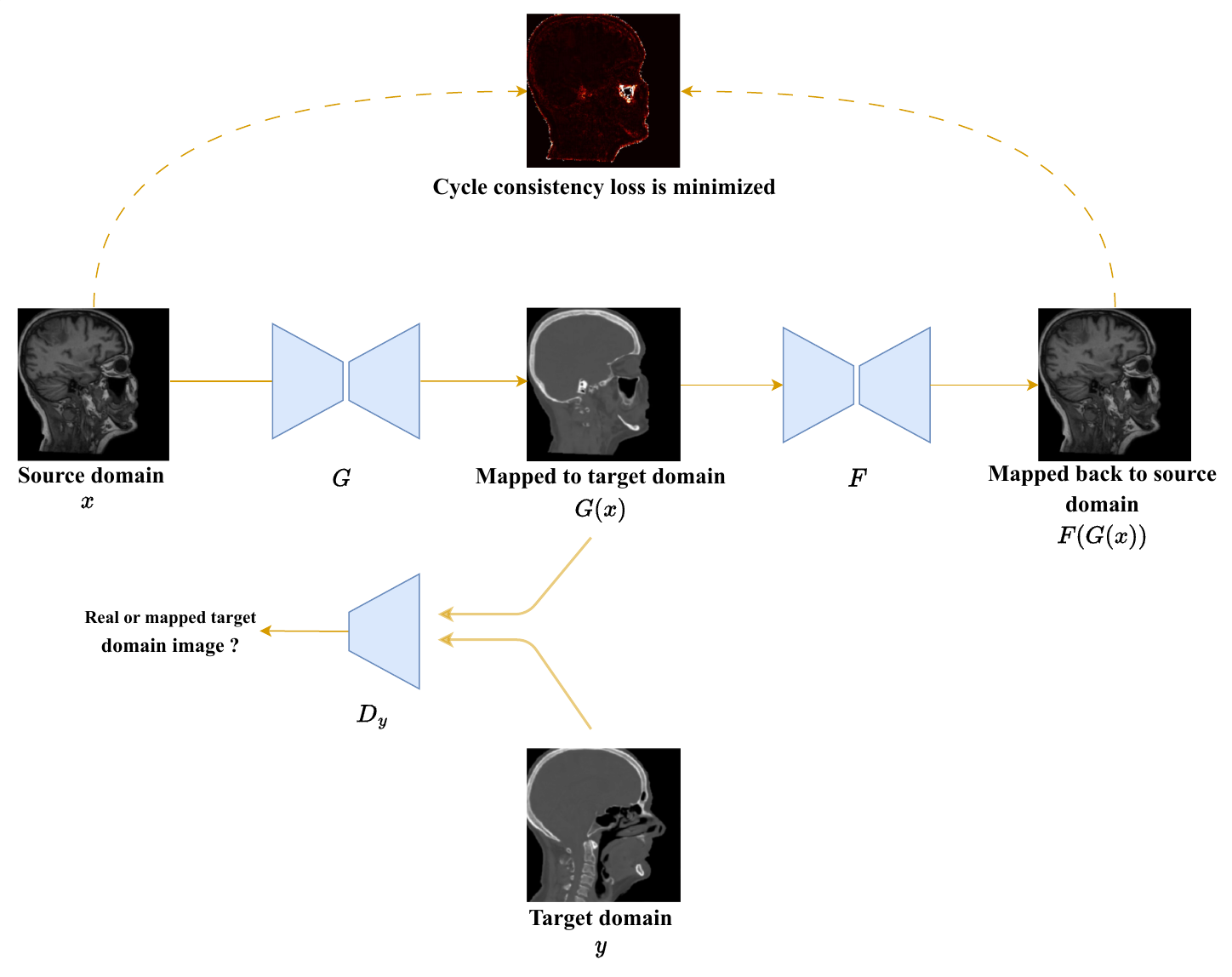}
\end{center}
\caption{Image-to-image translation via cycle consistency loss (Cycle-GAN) \citep{zhu2017unpaired}. The MR image from the source domain is converted to the target domain CT image and then converted back to the source domain. The objective is to minimize the discrepancy between the initial MR image and the reconstructed MR image.}  \label{fig_6}
\end{figure}

Image translation methods perform domain alignment not in feature space but in pixel space, converting source data to the ``style" of a target domain. GANs have been frequently employed in applications that require pixel-to-pixel mapping for image translation. CycleGAN \citep{zhu2017unpaired} is a common image-to-image (I2I) translation architecture that translates the features of one image domain into the other without any paired training samples. CycleGAN includes two mapping functions $G: X \rightarrow Y$ and $F: Y \rightarrow X$, and corresponding adversarial discriminators $D_Y$ and $D_X$ (see Fig.~\ref{fig_6}). $D_Y$ promotes $G$ to translate $X$ into results that are comparable to domain $Y$, and vice versa for $D_X$ and $F$. Two cycle consistency losses are further introduced to regularise the mappings and to capture the notion that going from one domain to another and back again should lead to the starting point: (a) forward cycle-consistency loss: $x\rightarrow G(x) \rightarrow F(G(x)) \approx x$, and (b) backward cycle-consistency loss: $y \rightarrow F(y) \rightarrow G(F(y)) \approx y$.

\textbf{Classification:} 
To classify breast cancer, Wollmann et al. \citep{wollmann2018adversarial} propose a CycleGAN-based domain adaptation approach. It involves transforming whole-slide images (WSIs) of lymph nodes from a source domain to a target domain using CycleGAN. Subsequently, a densely linked deep neural network (DenseNet) is utilized for the classification task. To address the limitations of cross-domain unpaired I2I translation methods that overlook class-specific semantics, Tang et al. \citep{tang2019tuna} introduce a task-driven CycleGAN framework. This framework is designed to preserve low-level details, high-level semantic information, and mid-level feature representation throughout the I2I translation process. An I2I translation method, noise adaptation GAN \citep{zhang2019noise}, focuses on modifying the noise patterns in test data to match those in the training data while preserving the image's content. This is achieved by treating noise as a style and employing two discriminators and one generator with four loss functions. Experiments are carried out for the classification of ultrasound images and the segmentation of OCT images. To tackle domain shift in multi-site imaging data, Robinson et al. \citep{robinson2020image} propose the use of image-and-spatial transformer networks (ISTNs), which were originally introduced by Lee et al. \citep{lee2019image}. The authors suggest utilizing adversarial training of ISTNs to achieve model-agnostic domain adaptation via explicit appearance and shape transformations between the different domains.

\textbf{Segmentation:} 
To perform the segmentation of White Matter Hyperintensity in Multicenter MR images, Palladino et al. \citep{palladino2020unsupervised} utilize the CycleGAN framework directly. This approach involves transforming the target domain images to have the same appearance as the source domain images before proceeding with the segmentation task on the target images. Additionally, Huo et al. \citep{huo2018synseg} propose an end-to-end framework that combines the cycle adversarial module with the segmentation module, enabling joint optimization of the two tasks for improved segmentation performance. Recently, Chen et al. \citep{chen2023segmentation} introduce a two-stage segmentation-guided domain adaptation network for multisite OCT data. This network is based on the CycleGAN architecture and incorporates novel composited loss function engineering. The goal is to achieve effective data harmonization during the domain adaptation process.

The CycleGAN structure used for I2I translation is not task-focused; it only serves the basic objective of appearance transfer. As a result, important prior knowledge such as organ borders, shapes, and appearance may not be adequately preserved during the synthesis process \citep{zhang2018task}. To address this limitation, several works \citep{zhang2018task, jiang2018tumor, chen2018semantic, jiang2020psigan} have introduced task-specific losses in the context of UDA. These task-specific losses are designed to improve the UDA process by providing additional constraints that align with the specific task requirements. By incorporating losses that are specific to the task, the training stability can be increased, and the risk of mode collapse, where the generator fails to produce diverse outputs, can be reduced \citep{bousmalis2017unsupervised}. A dense I2I translation network (DI2I) \citep{zhang2018task} incorporates a conditional GAN to preserve prior knowledge about the task, such as organ boundaries, shapes, and intensity variations. Finally, the image translation and segmentation are performed simultaneously in an end-to-end manner. In addition to the CycleGAN losses, Jiang et al. \citep{jiang2018tumor} introduce a tumor-aware loss to preserve the tumor structure during cross-domain adaptation and segmentation from CT to MRI. Similarly, Chen et al. \citep{chen2018semantic} enforce semantic consistency by aligning the predicted lung mask and the source domain ground truth mask for chest X-ray segmentation. In this approach, the segmentation model is trained independently from the domain adaptation GAN to provide more flexibility. For cross-modality MRI segmentation, Jiang et al. \citep{jiang2020psigan} set constraints to adequately control the appearance and geometry of structures of interest in I2I translation by integrating images and their segmentation probability maps as a joint density for adversarial learning. In \citep{zeng2021semantic}, the proposed model comprises an image translation module and a domain-specific segmentation module. The image translation module is based on a standard CycleGAN, whereas the segmentation module consists of two domain-specific segmentation networks: intra-modality semantic consistency (IMSC) and cross-modality semantic consistency (CMSC). The IMSC ensures that the reconstructed image after a cycle is segmented in the same manner as the original input image, while the CMSC encourages the synthesized images after translation to be segmented exactly the same as before translation. Both modules enforce semantic consistency within the network and yield impressive results in cardiac and hip segmentation tasks.

Some papers utilize the attention mechanism to capture long-range dependencies \citep{zhou2019cross, tomar2021self, kapil2021domain}. For multi-modal cardiac segmentation, Zhou et al. \citep{zhou2019cross} propose a cross-modal attention-guided Convolutional Network (CN). Initially, they employ CycleGAN for bidirectional image generation between MRI and CT modalities. Subsequently, they develop a novel CN that includes separate encoders for individual feature learning and a shared decoder for extracting shared features across modalities to ensure consistent segmentation. Finally, a cross-modal attention module is employed between the encoders and the decoder to leverage the correlated information between modalities. The entire network is then trained in an end-to-end manner. For cross-modality domain adaptation, Tomar et al. \citep{tomar2021self} utilize the dual cycle consistency loss to preserve semantic information during the image translation phase. They introduce a self-attentive spatial adaptive normalization approach consisting of two modules: the synthesis module and the attention module. The intermediate layers of the synthesis module receive the semantic layout information from the attention module, which facilitates learning the translation process. In their work, Kapil et al. \citep{kapil2021domain} modify the CycleGAN architecture with the following changes: first, they combine the input images and the segmentation mask; second, the first three convolutional layers of the two discriminators share weights between the prediction of the source distribution and the semantic segmentation posterior maps. They introduce self-attention blocks and spectral normalization to the discriminators and generators, respectively, to improve training stability and simulate long structural dependencies.

Du and Liu \citep{du2021constraint} propose a new constraint-based UDA network. This network enables mutual image translation between diverse domains, ensuring image-level domain invariance. Then, they feed the target domain into the source domain segmentation model to obtain pseudo-labels and employ cross-domain self-supervised learning between the two segmentation models. A new loss function is created to ensure the accuracy of the pseudo-labels. In addition, a cross-domain consistency loss is also introduced to improve the performance of whole-heart image segmentation.
In contrast to CycleGAN, Hu et al. \citep{hu2022domain} propose a multi-source UDA approach. They introduce an auxiliary high-frequency reconstruction task to facilitate UDA and mitigate the interference caused by replacing the low-frequency component. Additionally, they incorporate a domain-specific convolution module to enhance the segmentation model's ability to extract domain-invariant features.

\textbf{Others:} 
In their work, Xing et al. \citep{xing2019adversarial} introduce UDA for cell detection in cross-modality data. They utilize the CycleGAN framework to adapt the source images to the target domain and train a structured regression-based object detector using the adapted source images. Additionally, they fine-tune the detector using pseudo-labels from the target training data.
Building upon their previous work, Xing et al. \citep{xing2020bidirectional} extend the approach by incorporating bidirectional mapping. They perform I2I translation from source to target and target to source. They also extend this framework to the semi-supervised setting. In another extension of their work, Xing and Cornish \citep{xing2022low} not only address UDA for cell/nucleus detection but also tackle the challenge of sparse training target data.
Tierney et al. \citep{tierney2020domain} perform UDA for ultrasound beamforming, using simulated data as the source domain and real in vivo data as the target domain. A problem in this case is that domain shift occurs for both noisy input and clean output. They overcome this difficulty by expanding the CycleGAN network, where they use maps between synthetic simulation and actual in vivo domains to ensure the trained beamformers accurately represent the distribution of both noisy and clean in vivo data. The Annotation-free Network (ArcNet) \citep{li2022annotation} is used to recover images of the cataractous fundus. Cataract-like images are first simulated to create a source domain similar to the target domain. Then, a shared network with UDA is used to generalize the restoration model from the source domain to the target domain.

\begin{table*}[t]
    \centering
    \scriptsize
    \begin{center}
  %  \tabstyle{5pt}
    \caption{A joint approach combining feature alignment and image translation, which utilizes pixel, feature, and semantic information during adaptation while learning an invertible mapping through cycle consistency.}
    \label{tab:joint}
   \resizebox{\textwidth}{!}{
    \begin{tabular}{p{5cm}p{1.5cm}p{1.5cm}p{1.5cm}p{1.5cm}}
    \toprule
    & \multicolumn{1}{l}{Pixel Loss} & \multicolumn{1}{l}{Feature Loss} & \multicolumn{1}{l}{Semantic Loss} & \multicolumn{1}{l}{Cycle Consistency Loss} \\ \midrule 
Image translation methods (CycleGAN) & $\checkmark$ &  & & $\checkmark$ \\ 
Feature alignment  &  & $\checkmark$ & $\checkmark$ &  \\
Feature alignment + Image translation methods & $\checkmark$ & $\checkmark$ & $\checkmark$ & $\checkmark$ \\
\bottomrule

 \end{tabular}
}
\end{center}
\end{table*}

%%%%%%%%%%%%%%%%%%%%%%%%%%%%%%%%%%%%%%%%%%%%%%%%%%%%%%%%%%%%%%%%%%%%%%%%%%%%%%%%%%%%%%%%%%%

\subsection{Feature alignment + Image translation methods\label{subsec_3.3}} 
Feature alignment + image translation methods also called joint learning-based approaches involve two steps: first, image transformation alters the appearance of the source images so that they resemble the target domain; second, feature adaptation is used to close the remaining gap between the synthesized target-like images and the actual target images \citep{chen2020unsupervised}. The advantage of using a joint learning strategy is that it allows for the preservation of pixel, feature, and semantic information, which is not always achievable with individual alignment methods as shown in Table~\ref{tab:joint}.

\textbf{Segmentation:} 
The Cycle-Consistent Adversarial Domain Adaptation method (CyCADA), proposed by Hoffman et al. \citep{hoffman2018cycada}, is a joint learning approach designed for natural images. It consists of two phases: feature adaptation and image adaptation, which are trained sequentially in stages without direct interactions. CyCADA has been widely used as a base model in various medical imaging applications \citep{jia2019cone,liu2020unsupervised, liu2020pdam}. In detail, Jia et al. \citep{jia2019cone} employ image translation using CycleGAN to convert CT images to Cone-beam computed tomography (CBCT) images. To further mitigate the domain gap, they generate segmentation masks for both CBCT and converted CBCT images and incorporate them into a ``feature" discriminator. In their work, Liu et al. \citep{liu2020unsupervised} utilize a variant of CyCADA for instance segmentation, with Mask R-CNN as their baseline. Additionally, they perform semantic-level adaptation by considering the relationship between the foreground and background of nuclei images. They introduce a task re-weighting mechanism to restore the importance of each task loss. Furthermore, the authors enhance the approach proposed in \citep{liu2020unsupervised} by maximizing feature similarity \citep{liu2020pdam}.

In domain adaptation, feature alignment and image translation methods serve as complementary approaches at the feature and input levels, respectively. Hence, integrating both alignment aspects into a single framework can leverage their distinct advantages to enhance domain adaptation performance \citep{chen2019synergistic, chen2020unsupervised}. To handle the large domain gap in cross-modality segmentation tasks, unlike CyCADA, Chen et al. \citep{chen2019synergistic} propose a method called Synergistic Image and Feature Alignment (SIFA), which enables simultaneous feature and image translation. Specifically, the feature encoder is shared, allowing it to transform image appearance while extracting domain-invariant representations for the segmentation task. To further close the domain gap, the authors of \citep{chen2019synergistic} extend their work by incorporating the deep supervision technique into the feature alignment process and exploring bidirectional adaptation \citep{chen2020unsupervised}. The effectiveness of the proposed approach is demonstrated through experiments on cardiac substructure and abdominal multi-organ segmentation, yielding favorable results. Similarly, Han et al. \citep{han2021deep} introduce a unified joint learning framework that utilizes two symmetric translation sub-networks for bidirectional feature alignment. This framework incorporates all styled images to train the segmentation sub-network, leveraging adversarial loss and segmentation loss to effectively capture semantic information from diverse style images. To further improve the accuracy of domain adaptation, some studies leverage attention mechanisms in addition to image and feature alignment \citep{cui2021bidirectional, chen2022dual}. Specifically, Chen et al. \citep{chen2022dual} utilize an identical framework as SIFA. However, the feature space alignment of the framework is guided by the dual adversarial attention mechanism, focusing on particular regions indicated by the spatial and class attention mechanisms rather than treating all semantic feature elements equally. The proposed approach is evaluated on two tasks: skull segmentation based on MRI and cardiac substructure segmentation based on CT.

DSFN \citep{zou2020unsupervised} presents a dual-scheme fusion network for UDA. The network consists of two streams: one for source-to-target translation and the other for target-to-source translation. The data from these two streams are combined in a single network to make predictions. The proposed method is validated on the BRATS and MM-WHS datasets. U-Net GAN \citep{yan2019domain} addresses UDA for cardiac cine MRI images obtained from three different vendors. The method incorporates two loss functions: a structural similarity index loss at the image level which is a modified version of the CycleGAN loss, and a Unet loss at the feature level. Label-efficient UDA (LE-UDA) \citep{zhao2022uda} aims to address both domain shift and source label scarcity. The approach involves performing feature and image adaptation to mitigate the domain shift. Additionally, two teacher models are employed to leverage intra-domain and inter-domain information from multiple datasets to address source label scarcity. Dual Adaptive Pyramid Network (DAPNet) \citep{hou2019dual} addresses domain shift by utilizing multi-resolution feature-level and image-level losses. The image-level loss aims to reduce the color and style disparities between the two domains. On the other hand, the feature-level loss aims to minimize spatial inconsistencies between the two domains. To implement these losses, DAPNet employs adversarial training with domain classifiers. Recently, Li et al. \citep{li2023self} propose a self-training adversarial learning framework for UDA in retinal OCT fluid segmentation tasks. The framework incorporates joint image and feature alignment techniques.

%%%%%%%%%%%%%%%%%%%%%%%%%%%%%%%%%%%%%%%%%%%%%%%%%%%%%%%%%%%%%%%%%%%%%%%%%%%%%%%%%%%%%%%%%
 
 \subsection{Learning disentangled representations \label{subsec_3.4}}
Image translation methods, such as CycleGAN, typically assume a one-to-one mapping between the source and target domains. However, this assumption only partially captures the complex distribution of real-world data. In contrast, disentangled representations aim to achieve a many-to-many mapping by separating the representation into domain-specific features (DSFs) and domain-invariant features (DIFs). The core concept behind disentangled representation is to distinguish the content of an image from its style. The underlying assumption is that the content (anatomical information) remains consistent across domains, while the style is domain-specific (e.g., texture, lighting).

\begin{figure}[t]
\begin{center}
\includegraphics[scale=0.35]{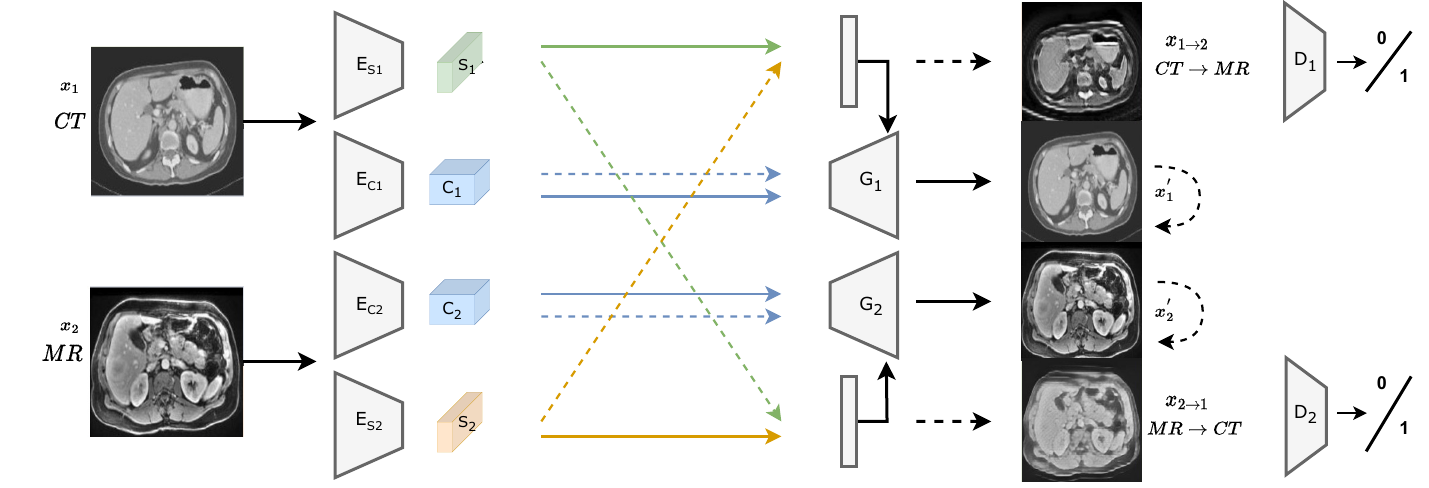}
\end{center}
\caption{Framework for disentangled representation learning module. Solid line: in-domain reconstruction, Dotted line: cross-domain translation. Adapted from Yang et al \citep{yang2019unsupervised}.}
\label{fig_7}
\end{figure}

\textbf{Segmentation:} 
To achieve disentangled representation \citep{yang2019unsupervised}, a style code $s_1$ and content code $c_1$ are initially extracted from the source image by encoder $E_{s_1}$ and $E_{c_1}$, respectively, in the same way, encoder $E_{s_2}$ and $E_{c_2}$ extract the respective style $s_2$ and content $c_2$ from the target domain. Now, a generator $G_1$ uses content code $c_1$ and style code $s_2$ to produce the image $t'$ in the target domain, just as generator $G_2$ employs content code $c_2$ and style code $s_1$ to produce image $s'$ in the source domain. The generators aim to deceive discriminators $D_1$ and $D_2$ by generating cross-domain images with switched style codes, as shown in Fig.~\ref{fig_7}. Further, the segmentation and image translation networks are trained independently for the liver segmentation task. In \citep{wang2022cycmis}, the approach is similar to \citep{yang2019unsupervised}, but with the integration of the segmentation stage and diverse image translation stage in an end-to-end manner, allowing the two parts to mutually benefit from each other. Furthermore, the network incorporates two semantic consistency losses to introduce regularization. For multi-domain image translation and multiple organ segmentation tasks, Jiang and Veeraraghavan \citep{jiang2020unified} propose a compact disentangled representation approach. They utilize a single universal content encoder and a VAE to extract image content and style codes from different domains. The VAE enforces the consistency of the style feature encoding with a global prior, assumed to cover all source and target modality styles.

Some works use the attention mechanism along with disentanglement to further alleviate the domain gap \citep{pei2021disentangle, sun2022attention}. In detail, Pei et al. \citep{pei2021disentangle} propose a method where the first step involves disentangling features into DIFs and DSFs. The encoder incorporates self-attention modules to enhance the representation of DIFs by allowing for long-range dependency modeling in image generation tasks. Additionally, the method minimizes a zero loss to ensure that the source (target) DSF information in target (source) images is as close to zero as possible. This ensures a clear separation of domain-specific information between the two domains. In \citep{sun2022attention}, the authors perform a coarse alignment step before disentanglement to address gaps such as brightness discrepancies between MRI and CT images. They then introduce an enhanced disentanglement strategy that utilizes the Hilbert-Schmidt independence criterion (HSIC) to promote independence and complementarity between DIFs and DSFs. Lastly, they employ attention bias to prioritize the alignment of task-relevant regions for cardiac segmentation. In \citep{shin2021unsupervised}, the authors employ a feature disentanglement method for small bowel segmentation. The method utilizes two encoders that independently extract intensity and non-intensity features. The non-intensity encoder takes the gradient images of the input image $X$, while the intensity encoder uses the actual image $X$. Two discriminators are utilized, one at the feature level and another at the output level, to improve alignment between the source and target domains.

Several works have jointly employed disentanglement and self-training \citep{xie2022unsupervised, yang2021mutual} to improve image segmentation accuracy. Specifically, Xie et al. \citep{xie2022unsupervised} employ two encoders: one for the domain-specific modality and another shared encoder for domain-invariant structures. They utilize a zero loss to enforce the domain-specific encoder to exclusively encode information from its own domain, thereby promoting the disentanglement of features. Additionally, self-training is employed to enhance the performance of the model. In \citep{yang2021mutual}, a coarse-to-fine prototype alignment is performed prior to the disentanglement of features for Polyp Segmentation. Furthermore, self-training is employed to further improve the model's performance. For cross-modality segmentation, Wang and Rui \citep{wang2022sgdr} propose a framework that utilizes four sets of encoders to disentangle features from original and generated images. The generated images are obtained by combining style and content codes from different modalities. In their work, Yao et al. \citep{yao2022novel} employ the disentanglement of features for 3D medical image segmentation, specifically focusing on maintaining semantic consistency at the depth level. They analyze the application of their research in the segmentation of multiple abdominal organs and brain structures. Li et al. \citep{li2022disentangled} perform the disentanglement of features for the purpose of cross-dataset subjective tinnitus diagnosis.

The GAN-based approach, named $MI^{2}GAN$ \citep{xie2020mi}, is designed to preserve image contents during I2I translation. In this approach, the content features and domain-specific information are separated for both the original and translated images. The decoupled content features are then utilized to maximize mutual information, aiming to retain the image objects and maintain their consistency. The suggested approach is evaluated through two tasks: the segmentation of OC and OD in fundus images, as well as polyp segmentation using colonoscopic images. Chen et al. \citep{chen2021beyond} extend on the work mentioned above by adopting an information bottleneck (IB) constraint to eliminate extraneous information, such as domain-specific information, while retaining the essential content features for preserving image objects. By incorporating the IB constraint, the authors reduce model complexity and eliminate the requirement for two decoders, as seen in the original $MI^{2}GAN$ framework.

%%%%%%%%%%%%%%%%%%%%%%%%%%%%%%%%%%%%%%%%%%%%%%%%%%%%%%%%%%%%%%%%%%%%%%%%%%%%%%%%%%%%%%%%

\subsection{Pseudo-labeling approach\label{subsec_3.5}}
Pseudo-labeling is a commonly employed technique in UDA to harness the potential of unlabeled data from the target domain. It entails assigning pseudo-labels to the unlabeled target domain data utilizing a model trained on the labeled source domain data. However, the pseudo-labels generated through this process are prone to noise due to the presence of domain gap. Therefore, a key aspect of pseudo-labeling is how different networks filter and remove the noise from the pseudo-labels to ensure their accuracy and reliability. 

\textbf{Classification:} 
For Epithelial-Stroma (ES) classification, Qi et al. \citep{qi2020curriculum} propose a curriculum learning approach. The authors begin by employing cross-domain similarity to identify reliable pseudo-labels for the target samples. Subsequently, they aim to reduce the intra-class distance by aligning the features of each class across both domains. Unlike traditional methods that rely on feature clusters, Qi et al. utilize centroids to mitigate any potential negative effects caused by the presence of pseudo-labels. This approach facilitates the improvement of ES classification accuracy and enhances the robustness of the model to domain shifts. 

\textbf{Segmentation:} 
For domain adaptive mitochondria segmentation, Wu et al. \citep{wu2021uncertainty} propose an uncertainty-aware model that incorporates Monte Carlo dropout layers into a U-Net backbone. The standard deviation of predictions obtained through dropout serves as a measure of uncertainty. For WMH segmentation, Strudel \citep{groger2021strudel} utilizes self-training and estimates the uncertainty of pseudo labels. The uncertainty information is incorporated into the training process using an uncertainty-guided loss function, which helps filter out labels with high certainty. In the context of UDA for instance segmentation of operating room (OR) images, Srivastav et al. \citep{srivastav2022unsupervised} employ natural images from the COCO dataset \citep{lin2014microsoft} as the source domain. They generate various augmentations of the target domain images to obtain accurate pseudo labels. Subsequently, these pseudo labels are used to train the model on OR images using a self-training framework. Additionally, they leverage disentangled feature normalization to mitigate the significant domain gap between the source and target images. 

\textbf{Others:}
For the image synthesis task, Liu et al. \citep{liu2021generative} propose the use of self-training for predicting continuous values as pseudo-labels. They employ an uncertainty mask to filter the pseudo-labels and leverage practical variational Bayes learning to measure the predictive confidence of the generated images. This approach enhances the quality and reliability of the synthesized images. In the context of cell detection, Cho et al. \citep{cho2021cell} utilize pseudo-cell-position heatmaps. They define a peak with a Gaussian distribution inside the heatmap to represent the cell centroid. However, the signal distribution around the peak often deviates from a Gaussian shape, even if the peak location is correct. To address this issue, the authors re-generate the pseudo-cell-position heatmap using the predicted peak positions, aiming to achieve a precise Gaussian shape. They also introduce a Bayesian network to select reliable pseudo-cell-position heatmaps. In their extended work, Cho et al. \citep{cho2022effective} incorporate curriculum learning into their approach. They start by selecting easier cases, allowing the model to learn from simpler instances, and gradually include more challenging cases step by step. This curriculum learning strategy further enhances the performance of cell detection. Mottaghi et al. \citep{mottaghi2022adaptation} propose a novel UDA approach to improve the efficacy of a surgical activity recognition model in a new operating room. They generate pseudo-labels for the target unlabeled videos. However, these pseudo-labels often exhibit an unbalanced distribution due to the model's higher confidence in dominant or simpler classes. To address the issue of unbalanced label distribution, the authors introduce a selection strategy for pseudo-labels. They utilize only a subset of pseudo-labels, determined by selecting the inverse of the frequency of each class. This selection approach ensures that the chosen pseudo-labels are more representative of infrequent or difficult classes, effectively mitigating the issue of unbalanced label distribution.

\begin{figure}[t]
\begin{center}
\includegraphics[scale=0.3]{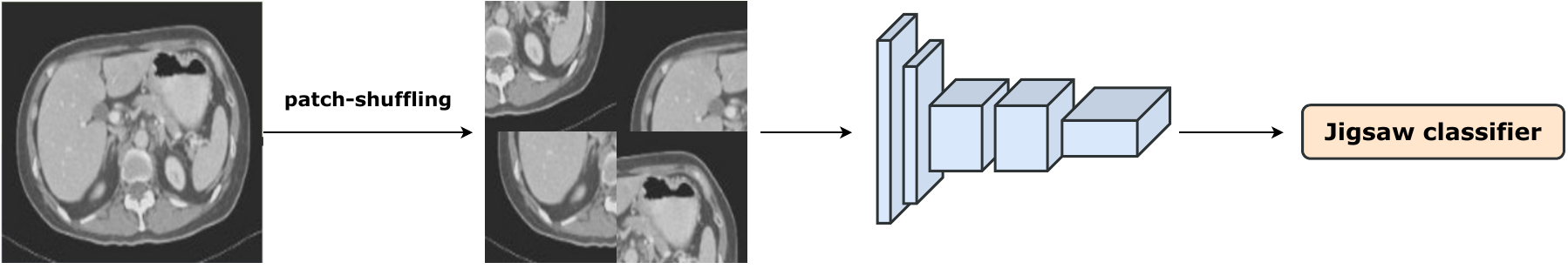}
\end{center}
\caption{A commonly used pretext task for self-supervised learning is solving a jigsaw puzzle.} 
\label{fig_8}
\end{figure}

\subsection{Self-supervision\label{subsec_3.6}}

Some works tackle UDA by employing the self-supervision approach. In this approach, alignment is achieved by simultaneously performing auxiliary self-supervised tasks on both domains. Each self-supervised task brings the two domains closer together along the direction relevant to that task. Training all the self-supervised tasks jointly with the primary task on the source domain has been shown to successfully generalize to the unlabeled target domain \citep{sun2019unsupervised}. There is a wide variety of auxiliary self-supervised tasks available. However, not all of them are applicable or suitable for UDA. Hence, the main challenge in the self-supervision method is to identify a suitable self-supervision task that allows the model to effectively learn useful representations from the data and induce alignment between the domains. 

\textbf{Classification:} For histology image classification, Koohbanani et al. \citep{koohbanani2021self} introduce Self-Path, a framework that addresses label scarcity and domain shift through a two-pronged approach, utilizing self-supervision. The framework is flexible and can incorporate various domain-specific or domain-agnostic pretext tasks and can be trained using adversarial or non-adversarial methods. Specifically, they introduce three novel domain-specific self-supervision tasks: predicting the magnification level, solving the magnification jigsaw puzzle (Fig.~\ref{fig_8}), and predicting the Hematoxylin channel. These tasks are designed to leverage contextual, multi-resolution, and semantic features present in histopathology images. 

\textbf{Segmentation:}
Some studies \citep{xue2020dual, cui2021structure} employ self-supervised learning to address UDA in cross-modality segmentation tasks. Specifically, Xue et al. \citep{xue2020dual} create an edge generation auxiliary task to support the primary segmentation task of the target domain, and hierarchical low-level adversarial learning is used to encourage low-level features of the domain to be uninformative in a hierarchical way. A structure-driven domain adaptation
approach \citep{cui2021structure} leverages an ordered set of 3D landmarks, which are representative points that reflect the common anatomical structure of the heart across different modalities. By learning to predict the positions of these landmarks, the model is able to capture and utilize the shared structural information. Furthermore, the high-level structural information is merged with an additional complementary feature, the Canny edges, to achieve effective results. For the segmentation of 3D multi-organ CT images acquired from different scanners/protocols, Fu et al. \citep{fu2020domain} create a jigsaw puzzle auxiliary task in which a CT scan is recovered from its shuffled patches to learn the spatial relationship between the images as shown in Fig.~\ref{fig_8}. A super-resolution network is also employed to standardize medical images from several domains to a specific spatial resolution. Finally, they train the auxiliary and super-resolution tasks alongside the organ segmentation task for effective performance. 

\textbf{Others:} 
Self-rule to multi-adapt (SRMA) \citep{abbet2022self} is utilized for cancer tissue detection. The approach leverages a small number of labeled source domain images and incorporates structural information from both domains by detecting visual similarities through intra-domain and cross-domain self-supervision. Furthermore, the model is extended to learn from multiple source domains, enhancing its adaptability to diverse datasets. Peng et al. \citep{peng2021vsr} apply DA to the area of 3D medical image super-resolution. They use in-plane and through-plane resolution distinctions of test data to achieve self-learned domain adaptation.

\begin{table*}[htbp]
    \centering
    \scriptsize
    \begin{center}
  %  \tabstyle{5pt}
    \caption{Overview of deep unsupervised domain adaptation methods.}
    \label{table:Unsupervised}
   \resizebox{\textwidth}{!}{
    \begin{tabular}{p{1cm}p{2.5cm}p{3cm}p{4.5cm}p{4cm}p{4.5cm}}
    \toprule
   Reference & \multicolumn{1}{l}{Domain-alignment method} & \multicolumn{1}{l}{Task} & \multicolumn{1}{l}{Dataset} & \multicolumn{1}{l}{Characterization of domain shift} & \multicolumn{1}{l}{Source domain $\rightarrow$ Target domain}\\ \midrule 
\multicolumn{2}{l}{\textbf{Feature Alignment Methods}} &&&& \\ \midrule

\citep{yu2022domain} & Feature (MMD) + Attention block & Classification: prediction of Subjective Cognitive
Decline & Alzheimer's disease neuroimaging initiative (ADNI) \citep{jack2008alzheimer} and Chinese Longitudinal Aging Study (CLAS) \citep{xiao2016china}  & Different scanners and scanning parameters in brain MRIs & ADNI $\rightarrow$ CLAS \\ \midrule

\citep{fang2023unsupervised} & Feature (MMD)& Major depressive order identification & REST-meta-MDD Consortium \citep{yan2019reduced} & Cross-domain rs-fMRI images  & Site/Hospital - 20 $\rightarrow$ Site/Hospital - 1\\ \midrule

\citep{sahu2020endo} & Consistency loss and supervised loss & Surgical tool segmentation in endoscopic video & Sim, SimRandom (SR), SimCholec (SC) \citep{pfeiffer2019generating} and 15
videos of the Cholec80 dataset (real) \citep{twinanda2016endonet} & Rendered and real datasets &Sim $\rightarrow$ real, SR $\rightarrow$ real, SC $\rightarrow$ real \\ \midrule

\citep{gomariz2022unsupervised} & Contrastive and supervised loss & Segmentation of retinal fluids in 3D OCT images & Two large OCT datasets \citep{maunz2021accuracy} & Acquired using two different imaging device (Spectralis and  Cirrus) & Spectralis $\rightarrow$ Cirrus \\ \midrule

\citep{wu2020cf} & CF distance & 1.Cardiac (LV and MYO) segmentation\newline 2.Cardiac (LV, RV, MYO) segmentation & 1.MM-WHS challenge dataset \citep{zhuang2016multi} \citep{zhuang2019evaluation} \newline 2.MS-CMRSeg challenge \citep{zhuang2018multivariate} & 1.Cross-modality whole heart dataset \newline 2.Multi modality MR dataset & 1.MRI $\rightarrow$ CT, CT $\rightarrow$ MRI \newline 2.BSSFP CMR $\rightarrow$ LGE CMR \\ \midrule

\citep{hu2022domain} & KL divergence & 3D Medical Image Synthesis & BraTS 2019 dataset \citep{menze2014multimodal}  & Multi modality MR images (2 subsets are used CBICA and TCIA) & CBICA $\rightarrow$ TCIA, TCIA $\rightarrow$ CBICA \\ \midrule

%5% fEATURE IMPLICIT

\citep{ren2018adversarial} &  Adversarial learning + Siamese architecture& Classification of prostate histopathology WSI’s & Cancer Genome Atlas (TCGA) dataset \citep{kandoth2013mutational} and local data set collected from Cancer Institute of New Jersey (CINJ)  &  Acquired from different institutes  & TCGA (31 institutes) $\rightarrow$ TCGA (1 institute), TCGA $\rightarrow$ CINJ\\ \midrule

\citep{zhang2019whole} & Adversarial learning + Entropy loss + Focal loss& Histopathology cancer classification & Private dataset & Cross-modality images & WSI $\rightarrow$ Microscopy images(MSIs)\\ \midrule

\citep{feng2023contrastive} & Conditional domain adversarial network + Contrastive loss & Automated pneumonia diagnosis & Radiological Society of North America (RSNA) dataset (Stage I) \citep{america2018rsna}, Child X-ray dataset \citep{kermany2018identifying}, RSNA+COVID and TTSH dataset & Acquired from different institutes  & RSNA dataset $\rightarrow$ Child X-ray dataset, RSNA+COVID $\rightarrow$ TTSH dataset \\ \midrule

\citep{cai2023prototype} & Multi-scale feature extraction module & Classification:  AD, MCI and CN identification & ADNI Database & Different scanners (3T and 1.5T MRI scanners) & 3T MRI scanner $\rightarrow$ 1.5T MRI scanner \\ \midrule

\citep{dou2018unsupervised, dou2019pnp}& Adversarial learning & Cardiac (AA, LA-blood, LV-blood, LV-myo) segmentation  & MM-WHS challenge dataset  & Cross-modality images  & MRI $\rightarrow$ CT, CT $\rightarrow$ MRI \\ \midrule

\citep{liu2021automated} & Adversarial learning + Attention mechanism & Cardiac (AA, LA-blood, LV-blood, LV-myo) segmentation  & MM-WHS challenge dataset & Cross-modality images & MRI $\rightarrow$ CT \\ \midrule

\citep{zeng2020entropy} & Feature and entropy-aligning discriminator & Hip cartilage segmentation & Private Dataset & Cross center images & MR (Center A) $\rightarrow$ MR (Center B) \\ \midrule

\citep{wang2019boundary} &Feature, entropy and boundary regions alignment discriminator& Segmentation of OD and OC in fundus images & Fundus dataset: Drishti-GS \citep{sivaswamy2014drishti}, REFUGE challenge dataset \citep{orlando2020refuge} and RIM-ONE-r3 \citep{fumero2011rim} & Acquired by different cameras & REFUGE $\rightarrow$ Drishti-GS, and RIM-ONE-r3 \\ \midrule

\citep{liu2019cfea} & Adversarial learning in Feature and output space & Segmentation of OD and OC in fundus images & REFUGE challenge dataset & Acquired by different cameras (color and texture) & Retinal fundus images (Zeiss Visucam 500 camera) $\rightarrow$ Retinal fundus images (Canon CR-2 camera) \\ \midrule

\citep{liu2021s} & Adversarial learning + Self-training & 1.Segmentation of OD and OC in fundus images \newline 2.Spinal Cord Gray Matter Segmentation  & 1.REFUGE challenge dataset, Drishti-GS (DGS) \newline 2.SCGM dataset & 1.Acquired by different cameras \newline 2. Multi-center and Multi-vendor & 1.REFUGE training set $\rightarrow$ DGS, REFUGE validation set \newline 2.SCGM(centers 1, 2, 3) $\rightarrow$ SCGM (center 4) \\ \midrule

 \citep{huang2022domain} & Adversarial learning + Pseudo labeling & Mitochondria segmentation & VNC III \citep{gerhard2013segmented}, lucchi [70], MitoEM-R \citep{wei2020mitoem}, MitoEM-H \citep{wei2020mitoem} & Different organism, tissue and camera & VNC III $\rightarrow$ Lucchi (Subset1), VNC III $\rightarrow$ Lucchi (Subset2), MitoEM-R $\rightarrow$ MitoEM-H, MitoEM-H $\rightarrow$ MitoEM-R \\ \midrule
 
\citep{li2022domain} & Feature, pseudo labeling & Nuclei instance segmentation and classification  & Lizard dataset (Large-Scale Dataset for Colonic Nuclear Instance Segmentation and Classification) \citep{graham2021lizard} & Different tissues i.e., digestive system tissues and colorectal adenocarcinoma gland tissues & Lizard (DigestPath) $\rightarrow$ Lizard (CRAG) \\ \midrule

\citep{bian2020uncertainty} & Feature, pseudo labeling & 1.Retinal and Choroidal layers of OCT images \newline 2.Cardiac (AA, LA-blood, LV-blood, LV-myo) segmentation & 1.Private dataset \newline 2.MM-WHS challenge dataset & 1.Cross-device dataset \newline
2.Cross-modality dataset & 1.OCT (Optovue device) $\rightarrow$ OCT (Heidelberg device) \newline 2.MRI $\rightarrow$ CT \\ \midrule

\citep{liu2022margin} & Contrastive loss + Pseudo labels + Entropy alignment by adversarial learning & Cardiac (AA, LA-blood, LV-blood, LV-myo) segmentation &  MM-WHS challenge dataset & Cross-modality images & MRI $\rightarrow$ CT, CT $\rightarrow$ MRI \\ \midrule

\citep{karaoglu2021adversarial} & Adversarial learning & Bronchoscopic Depth Estimation & Synthetic dataset and human pulmonary dataset \url{(https://www.dgiin.de/)} & Differences in illumination properties and anatomical discrepancy & Synthetic images $\rightarrow$ Real images \\ \midrule

\citep{yu2021cross} & Adversarial learning & Depth estimation in OCT Angiography & Private OCTA dataset & Cross-domain dataset & OCTA (CIRRUS HD-OCT 5000 system) $\rightarrow$ OCTA (RTVue XR Avanti system ) \\ \midrule

\citep{yoo2021transferring} & Adversarial learning + Two local discriminators - subject and stage &  Automated sleep staging & Montreal Archive of Sleep Studies (MASS) \citep{o2014montreal}, Sleep-EDF and Sleep-EDF-st database \citep{kemp2000analysis} \citep{goldberger2000physiobank} & Difference is the health conditions of the subjects & MASS database $\rightarrow$ Sleep-EDF database, Sleep-EDF-st database \\ \midrule

\citep{hao2022sparse} & Adversarial learning & OCTA Image Super-Resolution Reconstruction
& SUper-resolution REconstruction dataset (SURE) \citep{hao2022sparse} & Different imaging device and health conditions & SURE-O (Optovue imaging device) $\rightarrow$ SURE-Z (Zeiss imaging device) \\ \midrule

\citep{ouyang2019data} & VAE based feature prior matching + Adversarial training & Cardiac(LV-M, LA-B, LV-B, A-A) segmentation & MM-WHS challenge dataset (2017) & Cross-modality dataset & MRI $\rightarrow$ CT \\ \midrule

\citep{feng2022unsupervised} & Category-level regularization + Adversarial training & Fundus image segmentation &Fundus datasets: REFUGE challenge (training), RIMONE-r3 and Drishti-GS & Different fundus camera & REFUGE challenge $\rightarrow$ RIMONE-r3, REFUGE challenge $\rightarrow$ Drishti-GS \\ \midrule

% output space
\citep{wang2019patch} & Adversarial learning & Segmentation of OD and cup OC in fundus images & REFUGE challenge dataset, Drishti-GS dataset and RIM-ONE-r3 dataset & Acquired from different cameras so that the color and texture of the images are different & REFUGE Train $\rightarrow$ Drishti-GS, RIM-ONE-r3 , REFUGE Validation/Test \\ \midrule

\citep{haq2020adversarial} & Patch-based adversarial learning & Cell segmentation & Kidney Renal Clear cell carcinoma (KIRC) \citep{irshad2014crowdsourcing} WSIs and Triple Negative Breast Cancer (TNBC) histology images \citep{naylor2018segmentation} & Acquired from two different organs, institutions, imaging tools and protocols &  KIRC $\rightarrow$ TNBC, TNBC $\rightarrow$ KIRC \\ \midrule

%%Regularization based%%
\citep{wu2021unsupervised} & VAE + Regularization term & 1.Cardiac(RA, LA, MYO, RV and LV) segmentation \newline 2.Cardiac(MYO, RV and LV) segmentation & 1.MM-WHS challenge dataset \newline 2.MR images: MS-CMRSeg dataset (three CMR sequences, i.e.,the LGE , bSSFP and T2 images) & 1.Cross-modality dataset \newline 2.Multi modality dataset & 1.CT $\rightarrow$ MRI \newline 2.bSSFP images $\rightarrow$ LGE images \\ \bottomrule

\multicolumn{6}{r}{\footnotesize\textit{continued on the next page}}

\end{tabular}
}
\end{center}
\end{table*}

\begin{table*}[htbp]
\ContinuedFloat
    \centering
    \scriptsize
    \begin{center}
  %  \tabstyle{5pt}
    \caption{Overview of deep Unsupervised domain adaptation methods (continued)}
    
    \resizebox{\textwidth}{!}{
    \begin{tabular}{p{1cm}p{2.5cm}p{3cm}p{4.5cm}p{4cm}p{4.5cm}}
    \toprule
      Reference & \multicolumn{1}{l}{Domain-alignment method} & \multicolumn{1}{l}{Task} & \multicolumn{1}{l}{Dataset} & \multicolumn{1}{l}{Characterization of domain shift} & \multicolumn{1}{l}{Source domain $\rightarrow$ Target domain}\\ \midrule 

%% Image translation %%

%%5 GRAPH BASED %%

\citep{al2021olva} & VAE + OT theory & Cardiac(LV-M, LA-B, LV-B, A-A) segmentation & MM-WHS challenge dataset (2017) & Cross-modality dataset & MRI $\rightarrow$ CT \\ \midrule

\citep{lu2021learning} & VAE + Prior information (biomechanical standpoint) & Cardiac strain analysis & Cardiac images: Synthetic data and in vivo data \citep{lu2021learning} & Different appearance & Synthetic data $\rightarrow$ in vivo data \\ \midrule

\citep{liu2021prototypical} & Semantic-prototype interaction graph & Surgical instrument segmentation & Abdominal porcine procedure dataset: Endovis17 dataset \citep{allan20192017} and Endovis18 dataset \citep{allan20202018} & Recorded with different system & Endovis17 $\rightarrow$ Endovis18 \\ \midrule

\citep{liu2021graph} & Interactive Graph Network  & Instrument type, instrument part and binary segmentation & Abdominal porcine procedure dataset: Endovis17 dataset  and Endovis18 dataset & Recorded with different system  & Endovis17 + Endovis18 (training) $\rightarrow$ Endovis18 (test) \\ \midrule

\citep{chen2022graphskt} & Graph-Structured Knowledge Transfer & 1. Colonoscopic Polyp Detection \newline 2.Mammogram Mass Detection & 1.Colonoscopic images: ClinicDB \citep{bernal2015wm}, ETIS-LARIB \citep{silva2014toward}, Abnormal Symptoms in Endoscopic Images (ASEI) \citep{pogorelov2017nerthus}, In-house dataset (private) \newline
2.Mammogram datasets: DDSM \citep{Heath2007THEDD}, Inbreast \citep{moreira2012inbreast} & 1.Acquired from multiple sources \newline 2.Different source & 1.ClinicDB $\rightarrow$ ETIS-LARIB,  ClinicDB $\rightarrow$ ASEI, ClinicDB $\rightarrow$ In-house dataset \newline
2.Inbreast $\rightarrow$ DDSM \\ \midrule

\multicolumn{2}{l}{\textbf{Image Translation Methods}} &&&& \\ \midrule

\citep{wollmann2018adversarial} & CycleGan + Deep neural network (DenseNet) & Classification of WSIs and patient level breast cancer grading & CAMELYON17 dataset & Acquired from five medical centers & Source 2, 3, 4, 5 $\rightarrow$ source 1 \\ \midrule

\citep{tang2019tuna} & Task driven CycleGan & Pediatric pneumonia classification &
Adult chest X-ray dataset \url{(https: // www. kaggle. com/ c/ rsna-pneumonia-detection-challenge/ data)} and pediatric chest X-ray dataset \url{(https: // doi. org/ 10. 17632/ rscbjbr9sj. 3)} &  Different age & Adult dataset $\rightarrow$ Pediatric dataset \\ \midrule

\citep{zhang2019noise} &  Noise adaptation GAN & 1.OCT layer segmentation task \newline 2.Ultrasound image classification of bone contour integrity & 1.SINA \citep{chiu2012validated} dataset and private ATLANTIS retinal OCT dataset (AROD) \newline  2.Two datasets of vertebra ultrasound collected by SonixTABLET medical ultrasound system & 1.Collected using a different OCT machine \newline 2.Different setting (patients with thick muscles and thin muscles) & 1.AROD $\rightarrow$ SINA, SINA $\rightarrow$ AROD \newline 2. NA \\ \midrule

\citep{robinson2020image} & Image and spatial transformer networks &  Age regression and sex classification task & T1-weighted brain MRI from the Cambridge Centre for Ageing and Neuroscience study (Cam-CAN) \citep{taylor2017cambridge} and from the UK Biobank imaging study (UKBB) \citep{alfaro2018image} & Acquisition and Population differences & Cam-CAN (UKBB) $\rightarrow$ UKBB (Cam-CAN) \\ \midrule

%% Image translation segmentat%%

 \citep{palladino2020unsupervised} & CycleGan & White Matter Hyperintensity Segmentation & MICCAI  white matter hyperintensities WMH Challenge \citep{kuijf2019standardized} & Multi-center(Utrecht, Singapore, Amsterdam) and different scanner image database & Utrecht, Singapore, Amsterdam $\rightarrow$ Utrecht, Singapore, Amsterdam \\ \midrule

\citep{huo2018synseg} & CycleGan and segmentation network in an end-to-end framework & 1.MRI to CT splenomegaly synthetic segmentation 2.CT to MRI TICV synthetic segmentation & Different public and private datasets & Cross-modality dataset & MRI $\rightarrow$ CT, CT $\rightarrow$ MRI \\ \midrule

\citep{chen2023segmentation} & A two-staged CycleGAN-based network & Layer segmentation of retinal OCT images & Zeiss Cirrus 5000 HD-OCT machine and Topcon 3D OCT 1000 Mk2 machine (UK Biobank database \citep{sudlow2015uk}) & Different acquisition machines and organizations & Zeiss $\rightarrow$ Topcon \\ \midrule

\citep{zhang2018task} & CycleGan + Prior knowledge about the task & Multi-organ (lung, heart, liver, bone) segmentation & Digitally Reconstructed Radiographs (synthetic data re generated from labeled CTs) and real X-ray images & Different appearance & Digitally Reconstructed Radiographs (DRRs) $\rightarrow$ X-ray \\ \midrule

\citep{jiang2018tumor} & CycleGan + Tumor-aware loss & Lung Cancer Segmentation
& The Cancer Imaging Archive (TCIA) CT dataset \citep{clark2013cancer} and Private MRI dataset & Cross-modality dataset & CT $\rightarrow$ MRI \\ \midrule

\citep{chen2018semantic} & CycleGan + Segmentation mask alignment & Lung segmentation (right lung and left lung) & Chest x-ray datasets: Montgomery set \citep{jaeger2014two} and JSRT set \citep{shiraishi2000development} & Different disease type, intensity, and contrast & Montgomery set $\rightarrow$ JSRT set \\ \midrule

\citep{zhou2019cross} & CycleGan + Cross-modal attention & Cardiac (RVC, LAC, LVC, RAC, MYO, AA and PA) segmentation & MM-WHS challenge dataset & Cross-modality dataset & MRI $\rightarrow$ CT, CT $\rightarrow$ MRI \\ \midrule

\citep{tomar2021self} & Dual CycleGan + Self-attentive spatial adaptive normalization & 1.Brain tumor segmentation \newline 2.Cardiac(AA, LA-blood, LV-blood, LV-myo) & 1.BraTS\newline 2.MM-WHS challenge dataset &  1.Multimodal brain tumor MRI (T1/T2) \newline
2. Cross-modality whole heart &  1.MRI-T2 $\rightarrow$  MRI-T1,  MRI-T1 $\rightarrow$ MRI-T2\newline 2. MRI $\rightarrow$ CT, CT $\rightarrow$ MRI \\ \midrule

\citep{kapil2021domain} & CycleGan + Self-attention blocks and spectral normalization & Epithelial Segmentation, Tumor Cell Scoring and survival Analysis & Private Pan-Cytokeratin  (PanCK) stained WSIs and programmed death ligand1 (PD-L1) clone WSIs & The PanCK images and the PD-L1 images are unpaired and come from two independent patient cohort & PanCK $\rightarrow$ PD-L1, PD-L1 $\rightarrow$ PanCK \\ \midrule

\citep{du2021constraint} & CycleGan + Pseudo labels &
Whole heart segmentation( (AA, LA-blood, LV-blood, LV-myo) & MM-WHS challenge dataset & Cross-modality dataset & MRI $\rightarrow$ CT \\ \midrule

\citep{hu2022domain} & High Frequency Reconstruction & Joint OC and OD segmentation & RIGA+ dataset & Multi-domain fundus image dataset 
& BinRushed, Magrabia $\rightarrow$ MESSIDOR-BASE1, MESSIDOR-BASE2, MESSIDOR-BASE3 \\ \midrule

%% Image translation Others%%

\citep{xing2019adversarial}  & CycleGan + Pseudo labels  & Cell/nucleus detection task & Cancer histology images stained with hematoxylin and eosin (H\&E) \citep{sirinukunwattana2016locality}, pancreatic neuroendocrine tumor bright-field microscopy images stained with immunohistochemical (IHC) \citep{xing2019pixel} Ki67 and cervical cancer HeLa cell line images generated with phase-contrast (PC) \citep{arteta2012learning} microscopy & Different staining techniques or imaging protocols/modalities & H\&E $\rightarrow$ PC, H\&E $\rightarrow$ IHC \\ \midrule

\citep{xing2020bidirectional} & Bidirectional image alignment & Cell/nucleus detection task & H\&E, IHC, PC and DAPI stained
fluorescence microscopy image collection \citep{tofighi2019prior} & Different staining techniques or imaging protocols/modalities & H\&E, IHC, DAPI $\rightarrow$ H\&E, IHC, DAPI \\ \midrule

\citep{li2022annotation} & Image translation & Restoration of cataract fundus images & RCF (private), DRIVE, Kaggle \url{(https://www.kaggle.com/datasets/jr2ngb/cataractdataset)} &
Different health conditions & DRIVE, cataractous subset of kaggle $\rightarrow$ RCF, normal subset of kaggle \\ \midrule

\multicolumn{2}{l}{\textbf{Feature Alignment + Image Translation Methods}} &&&& \\ \midrule

\citep{chen2019synergistic} & Synergistic image and feature alignment (SIFA) & Cardiac (AA, LA-blood, LV-blood, LV-myo) & MM-WHS challenge dataset & Cross-modality dataset & MRI $\rightarrow$ CT \\ \midrule

\citep{chen2020unsupervised} & SIFA and Deep supervision technique into the feature alignment & 1.Cardiac (AA, LA-blood, LV-blood, LV-myo) \newline 2.Multi organ segmentation & 1.MM-WHS challenge dataset \newline 2.T2-SPIR MRI training data from the ISBI 2019 CHAOS (Combined Healthy Abdominal Organ Segmentation) Challenge \citep{kavur2019chaos} and public CT data \citep{landman2015multi} & 1.Cross-modality dataset \newline 2.Cross-modality dataset & 1. MRI (CT) $\rightarrow$  CT (MRI) \newline
2. MRI (CT) $\rightarrow$  CT (MRI) \\ \midrule

\citep{jia2019cone} & Image and feature (mask) alignment  & Cone-beam computed tomography (CBCT) segmentation & Private Dataset & Cross-modality dataset & CT images $\rightarrow$ CBCT images \\ \bottomrule

\multicolumn{6}{r}{\footnotesize\textit{continued on the next page}}\\

\end{tabular}
}
\end{center}
\end{table*}

\begin{table*}[htbp]
\ContinuedFloat
    \centering
    \scriptsize
    \begin{center}
  %  \tabstyle{5pt}
    \caption{Overview of deep Unsupervised domain adaptation methods (continued)}
   \resizebox{\textwidth}{!}{
    \begin{tabular}{p{1cm}p{2.5cm}p{3cm}p{4.5cm}p{4cm}p{4.5cm}}
    \toprule
    Reference & \multicolumn{1}{l}{Domain-alignment method} & \multicolumn{1}{l}{Task} & \multicolumn{1}{l}{Dataset} & \multicolumn{1}{l}{Characterization of domain shift} & \multicolumn{1}{l}{Source domain $\rightarrow$ Target domain}\\ \midrule

\citep{liu2020unsupervised} & Nuclei inpainting mechanism and Semantic alignment & Nuclei instance segmentation & Histopathology datasets (kumar \citep{kumar2017dataset}, TNBC \citep{naylor2018segmentation}) fluorescence microscopy dataset(BBBC039V1 \citep{ljosa2012annotated}) & Different diseases and devices & BBBC039V1 $\rightarrow$ kumar, TNBC \\ \midrule

\citep{liu2020pdam} & Nuclei inpainting mechanism, Semantic alignment and maximises feature similarity  & 1.Nuclei instance segmentation \newline 2.Instance mitochondria segmentation & 1.Histopathology datasets (kumar, TNBC) fluorescence microscopy dataset(BBBC039V1) \newline
2.EPFL dataset \citep{lucchi2013learning} obtained from the mouse brain hippocampus and VNC dataset taken from Ventral Nerve Cord \citep{gerhard2013segmented} &  1.Different diseases and devices \newline 2.Different diseases and devices &  1. BBBC039V1 $\rightarrow$  kumar, TNBC \newline 2. EPFL dataset $\rightarrow$ VNC dataset \\ \midrule

\citep{han2021deep} & CycleGan + Bidirectional feature alignment &  1.Cardiac (AA, LA-blood, LV-blood, LV-myo) segmentation \newline 2.Brain Tumor Segmentation & 1.MM-WHS challenge dataset \newline 2.BRATS dataset  & 1.Cross-modality dataset \newline 2.Multi-modality dataset & 1.MRI (CT) $\rightarrow$  CT (MRI) \newline 2. T2 $\rightarrow$ FLAIR, T1, T1CE \\ \midrule

\citep{chen2022dual} & CycleGan + Feature space alignment is led by the dual adversarial attention mechanism & 1.Skull segmentation \newline 2.Cardiac (AA, LA-blood, LV-blood, LV-myo) segmentation & 1.CT images from CQ500 Database \citep{chilamkurthy2018deep} and Alzheimer’s Disease Neuroimaging Initiative (ADNI) database (MRI) \citep{trzepacz2014comparison} \newline 2.MM-WHS challenge dataset & 1.Cross-modality \newline 2.Cross-modality dataset & 1.CT $\rightarrow$ MRI 2.MRI $\rightarrow$ CT \\ \midrule

\citep{zou2020unsupervised} & Dual-scheme fusion network & 1.Cardiac (AA, LA-blood, LV-blood, LV-myo) segmentation \newline 2.Brain Tumor Segmentation & 1.MM-WHS challenge dataset \newline 2.BRATS dataset & 1.Cross-modality dataset \newline 2.Multi - modality dataset & 1.MRI $\rightarrow$ CT  \newline 2.T2 $\rightarrow$ FLAIR (fluid-attenuated inversion recovery), T1, T1CE \\ \midrule

\citep{yan2019domain} & Modified version of CycleGan and Unet loss at feature level & LV segmentation  & Private MRI images & MRI images from 3 MRI machines (44 Philips samples, 50 GE samples, 50 Siemens samples) & Siemens $\rightarrow$ Philips, GE $\rightarrow$ Philips \\ \midrule

\citep{zhao2022uda} & CycleGan and feature alignment & 1.Abdominal multi-Organ Segmentation (liver, right kidney, left kidney, and spleen) \newline 2.Cardiac (AA, LA-blood, LV-blood, LV-myo) segmentation & 
1.MICCAI 2015 Multi-Atlas Abdomen Labeling Challenge (CT images), ISBI 2019 CHAOS Challenge (MR images) \newline 2.MM-WHS challenge dataset & 1.Cross-modality dataset \newline 2.Cross-modality dataset &
1.MRI (CT) $\rightarrow$ CT (MRI) \newline 2.MRI (CT) $\rightarrow$ CT (MRI) \\ \midrule

\citep{hou2019dual} & CycleGan and feature alignment & Cross-Stain Histopathology Image Segmentation & Warwick-QU dataset \citep{sirinukunwattana2017gland} (WSIs) stained with H\&E and GlandVision dataset \citep{fu2012glandvision} contains DABH stained colon images & Different staining mechanism and labeling strategy & Warwick-QU $\rightarrow$ GlandVision, GlandVision $\rightarrow$ Warwick-QU \\ \midrule

\multicolumn{2}{l}{\textbf{Learning Disentangled Representations}} &&&& \\ \midrule

\citep{yang2019unsupervised} & Disentangled Representation & Liver segmentation & LiTS challenge 2017 dataset \citep{christ2017lits} (CT slices) and multi-phasic (MRI slices) & Cross-modality dataset  & CT images $\rightarrow$ MR images \\ \midrule

\citep{wang2022cycmis} & Disentangled Representation + Semantic consistency loss  & 1.Cardiac (LV and MYO) segmentation \newline 2.Cardiac (LV, MYO and RV) & 1.Multi-Sequence Cardiac MR Segmentation (MS-CMRSeg) \newline 2.MM-WHS challenge dataset(2019) & 1.Different MR imaging protocols \newline 2. Cross-modality & 1.bSSFP CMR $\rightarrow$ LGE CMR images \newline 2. MRI (CT) $\rightarrow$  CT (MRI) \\ \midrule

\citep{jiang2020unified} & Compact disentangled Representation & Multi-organ segmentation &
CHAOS (MR) challenge data and public CT data & Cross-modality dataset & CT $\rightarrow$ T1-weighted and T2-weighted MRI \\ \midrule

\citep{pei2021disentangle} & Disentangled Representation + Self attention modules & 1.LV, MYO and RV segmentation \newline 2. LV, MYO and RV segmentation & 1.MS-CMRSeg dataset \newline 2.MM-WHS challenge dataset (2019) & 1.Different MR imaging protocols \newline 2.Cross-modality dataset & 1.bSSFP CMR $\rightarrow$ LGE CMR images \newline 2.MRI (CT) $\rightarrow$  CT (MRI) \\ \midrule

\citep{sun2022attention} & Disentangled Representation + HSIC + Attention bias  & AA, MYO, RV and LV segmentation & MMWHS challenge 2017 dataset & Cross-modality dataset & MRI (CT) $\rightarrow$  CT (MRI) \\ \midrule

\citep{shin2021unsupervised} & Disentangled Representation + Feature and output alignment
& Small Bowel Segmentation & Private contrast-enhanced abdominal CT scans & Dataset is composed of two subsets, first subset acquired with oral contrast and second subset without any oral contrast & First subset acquired with oral contrast $\rightarrow$ Second subset without any oral contrast \\ \midrule

\citep{xie2022unsupervised} & Disentangled Representation + Self training & 1.Cardiac(AA, MYO, RV and LV ) \newline 2.Multi-organ (liver, right kidney, left kidney, and spleen) \newline
3.Brain tumor segmentation & 1.MMWHS challenge 2017 dataset \newline
2.ISBI 2019 CHAOS Challenge (MR images) and CT data from \citep{landman2015multi} \newline
3.BRATS dataset & 1.Cross-modality \newline 2.Cross-modality \newline 3.Multi-modality dataset & 1.MRI (CT) $\rightarrow$  CT (MRI) \newline 2.MRI (CT) $\rightarrow$  CT (MRI) \newline 3.T2 (Flair) $\rightarrow$ Flair (T2) \\ \midrule

\citep{yang2021mutual} & Disentangled Representation + Self training & Polyp Segmentation & Colonoscopic images: CVC-DB \citep{bernal2012towards}, Kvasir-SEG \citep{jha2020kvasir} and ETIS-Larib \citep{naylor2018segmentation} & Captured by different centers using different colonoscopy devices & CVC-DB, Kvasir-SEG $\rightarrow$  ETIS-Larib \\ \midrule

\citep{wang2022sgdr} & Disentangled Representation &1.LV, MYO and RV segmentation \newline 2.LV, MYO and RV
& 1.MM-WHS challenge dataset (2019) \newline 2.MS-CMRSeg dataset & 1.Cross-modality dataset \newline 2.Different MR imaging protocols & 1.MRI (CT) $\rightarrow$ CT (MRI) \newline 2.bSSFP CMR $\rightarrow$  LGE CMR images \\ \midrule

\citep{yao2022novel} & Disentangled Representation &1.Cardiac structures (cochlea and vestibular schwannoma) 2.Multi-organ (liver, right kidney, left kidney, and spleen) segmentation & 1.Vestibular schwannoma segmentation dataset \newline
2.ISBI 2019 CHAOS Challenge (MR images) and CT data from \citep{landman2015multi} & 1.Contrast-enhanced T1 (ceT1) and high-resolution T2 (hrT2) MRI \newline 2.Cross-modality & 1.ceT1 $\rightarrow$  hrT2 \newline 2.MRI (CT) $\rightarrow$   CT (MRI) \\ \midrule

\citep{xie2020mi} & Disentangled Representation + Mutual information & 1.Polyp segmentation on colonoscopy images \newline 2.OD and OC segmentation & 1.Colonoscopic images: CVC-Clinic \citep{vazquez2017benchmark} and ETIS-Larib \newline 2.REFUGE challenge dataset & 1.Multi-centre dataset \newline
2.Two different fundus camera & 1.CVC-Clinic dataset $\rightarrow$ ETIS-Larib \newline 2.REFUGE(training) $\rightarrow$  REFUGE (validation) \\ \midrule

\citep{chen2021beyond} & Disentangled Representation + Mutual information + Information bottlenek & 1.polyp segmentation on colonoscopy images \newline
2.OD and OC segmentation \newline 3.cardiac(AA, MYO, RV and LV) & 1.CVC-Clinic and ETIS-Larib \newline  2.REFUGE challenge dataset \newline 3.MMWHS challenge 2017 dataset & 1.Multicentre dataset \newline 2.Different fundus camera \newline 3.Cross-modality & 1.CVC-Clinic dataset $\rightarrow$ ETIS-Larib \newline 2.REFUGE(training) $\rightarrow$ REFUGE (validation) \newline 3.MRI $\rightarrow$ CT \\ \midrule

\multicolumn{2}{l}{\textbf{Pseudo - Labeling Methods}} &&&& \\ \midrule

\citep{qi2020curriculum}& Curriculum learning & Epithelial-stroma (ES) classification &
Breast cancer tissue samples from Netherland Cancer Institute's (NKI) dataset and Vancouver General Hospital's (VGH) \citep{beck2011systematic} and colorectal cancer tissue samples from IHC dataset \citep{linder2012identification} & Different specific diseases, staining methods and imaging scanners  & NKI, VGH $\rightarrow$ IHC, VGH, IHC $\rightarrow$ NKI, NKI, IHC $\rightarrow$ VGH \\ \midrule

\citep{groger2021strudel} & Self training + Uncertainty-guided loss & White Matter Hyperintensity Segmentation & WMH challenge dataset and Alzheimer’s Disease Neuroimaging Initiative (ADNI) \citep{beckett2015alzheimer} & Different source & WMH $\rightarrow$ ADNI-2 dataset \\ \bottomrule

\multicolumn{6}{r}{\footnotesize\textit{continued on the next page}}\\

\end{tabular}
}
\end{center}
\end{table*}

\begin{table*}[t]
\ContinuedFloat
    \centering
    \scriptsize
    \begin{center}
  %  \tabstyle{5pt}
    \caption{Overview of deep Unsupervised domain adaptation methods (continued)}
    
    \resizebox{\textwidth}{!}{
    \begin{tabular}{p{1cm}p{2.5cm}p{3cm}p{4.5cm}p{4cm}p{4.5cm}}\toprule
    Reference & \multicolumn{1}{l}{Domain-alignment method} & \multicolumn{1}{l}{Task} & \multicolumn{1}{l}{Dataset} & \multicolumn{1}{l}{Characterization of domain shift} & \multicolumn{1}{l}{Source domain $\rightarrow$ Target domain}\\ \midrule 

\citep{srivastav2022unsupervised} & Self training + Augmentation & Instance segmentation in the operating room & COCO \citep{lin2014microsoft} and two OR datasets MVOR \citep{srivastav2018mvor} and TUM-OR \citep{belagiannis2016parsing} & Cross domain dataset & COCO $\rightarrow$ MVOR, TUM-OR \\ \midrule

\citep{liu2021generative} & Self training + Uncertainty mask + Variational Bayes learning & tagged-to-cine MR image synthesis task & Private dataset & 1.Cross-scanner \newline 2.Cross-center & 1.Clinical center A (scanner 1) $\rightarrow$ Clinical center A (scanner 2) \newline
2.Clinical center A $\rightarrow$ Clinical center B \\ \midrule

\citep{cho2021cell} & Pseudo-cell-position heatmap + Bayesian network & Cell Detection & 1.C2C12 dataset \citep{ker2018phase} cultured under 3 conditions Control: (no growth factor),  FGF2 (fibroblast growth factor) and BMP2 \newline 2.Human mammary epithelial cells (HMEC) dataset \citep{nieto2016emt} cultured under 2 conditions Control (no stimulus) and EMT (epithelial-mesenchymal transition) &
1.Different cultured conditions (appearance \newline 2. Different cultured conditions  & 1.control,  FGF2, BMP2 $\rightarrow$ control,  FGF2, BMP2 \newline 2.Control, EMT $\rightarrow$ EMT, Control \\ \midrule

\citep{cho2022effective} & Pseudo-cell-position heatmap + Bayesian network + Curriculum learning & Cell Detection & 1.C2C12 dataset \citep{ker2018phase} cultured under 3 conditions Control: FGF2, BMP2 and BMP2+FGF2 \newline 2.HMEC dataset cultured under 2 conditions Control and EMT & 1.Different cultured conditions (appearance) \newline 2.Different cultured conditions & 1.Control,  FGF2, BMP2, FGF2 + BMP2 $\rightarrow$ Control, FGF2, BMP2, FGF2 + BMP2 \newline 2.Control, EMT $\rightarrow$ EMT, Control \\ \midrule

\citep{mottaghi2022adaptation} & Pseudo labeling & Surgical activity recognition models across operating rooms & Dataset of full-length surgery videos from two robotic ORs (OR1 and OR2) & Distinct layout and type of procedures and teams & OR1 (OR2) $\rightarrow$ OR2 (OR1) \\ \midrule

\multicolumn{2}{l}{\textbf{Self-Supervision}} &&&& \\ \midrule

\citep{koohbanani2021self} & Prediction of magnification level and Hematoxylin channel + Solving jigsaw puzzle & Classification of Pathology Images & WSIs: Camelyon16 dataset[footnote] and In house dataset (LNM-OSCC) & Different center and scanner & Camelyon16 dataset $\rightarrow$ LNM-OSCC dataset \\ \midrule

\citep{xue2020dual} &  Edge generation task + Adversarial learning & Cardiac(AA, MYO, RV and LV) segementation & MM-WHS challenge dataset (2017) & Cross-modality dataset & MRI $\rightarrow$ CT \\ \midrule

\citep{cui2021structure} & SSL of an ordered set of 3D landmarks + Canny edges & Cardiac(AA, MYO, RV and LV) segmentation & MM-WHS challenge dataset (2017) & Cross-modality dataset & MRI (CT) $\rightarrow$ CT (MRI) \\ \midrule

\citep{fu2020domain} & Jigsaw puzzle auxiliary task + Super-resolution network & 8 abdominal organs, including Aorta, Gallbladder, Left Kidney, Right Kidney, Liver, Pancreas, Spleen and Stomach & Private multi organ dataset, Synapse dataset \url{(https://www.synapse.org/\#!Synapse:syn3193805/wiki/217789)}, 4 single-
organ datasets \citep{simpson2019large} and NIH Pancreas dataset \url{(https://wiki.cancerimagingarchive.net/display/Public/Pancreas-C)} & Different organization & Private multi organ dataset $\rightarrow$ Synapse dataset, NIH Pancreas dataset, 4 single-organ datasets \\ \midrule

\citep{abbet2022self} & Intra-domain and cross-domain self-supervision & 1.Colorectal tissue type classification \newline2.Multi-source patch classification & Kather-16 \citep{kather2016multi}, Kather-19 \citep{kather2019predicting}, Colorectal cancer tissue phenotyping dataset (CRC-TP) \citep{javed2020cellular} and In-house dataset & Discrepancies in class definitions between datasets & 1.Kather-19 $\rightarrow$ Kather-16 , Kather-19 $\rightarrow$ In-house dataset \newline 2.Kather-19 + Kather-16 $\rightarrow$ CRC-TP \\ \midrule

\citep{peng2021vsr} & Self-supervision & Medical image super-resolution (SR) & LIDC-IDRI \citep{armato2011lung} dataset (CT scans) and private CT datasets that are acquired for different organs & Different organs & Training - 810 volumes of LIDC-IDRI dataset,Testing - 50 volumes of LIDC-IDRI dataset and private CT datasets \\ \bottomrule
\end{tabular}
}
\end{center}
\end{table*}

%%%%%%%%%%%%%%%%%%%%%%%%%%%%%%%%%%%%%%%%%%%%%%%%%%%%%%%%%%%%%%%%%%%%%%%%%%%%%%%%%%%%%%%%%%%%

\section{Emerging areas}
\label{sec_4}
\subsection{Source free unsupervised domain adaptation (SFUDA)}

Existing studies on deep learning for UDA heavily rely on the availability of source data. However, accessing source data is often challenging in practical scenarios due to reasons such as data privacy protection. Therefore, there is a growing demand for source-free UDA (SFUDA) methods that can transfer a pre-trained source model to an unlabeled target domain without requiring access to any source data. In the following sections, we discuss different methods used for SFUDA in medical imaging.

\textbf{Data generation approach:}
Several research works focus on generating image data similar to the source domain and accomplishing cross-domain adaptation by utilizing standard UDA methods. In deep learning models, batch normalization (BN) maintains the running mean and variance for a mini-batch of training data within each layer. Some studies explicitly leverage BN statistics for SFUDA \citep{yang2022source, hong2022source}. In detail, Yang et al. \citep{yang2022source} propose a two-stage approach for generating source-like images in a coarse-to-fine manner. In the first stage, they leverage the batch normalization statistics saved in the source model to retain the style features of source images while preserving the content information of the target data. The second stage involves a fine image generation step, where an image generator based on Fourier Transform is used to enhance image quality by removing ambiguous textural components from the generated images. To facilitate adaptation, the authors utilize the generated source-like images alongside the target images. They design a contrast distillation module for feature-level adaptation and a compact consistency measurement module for output-level adaptation. These modules enable the alignment of features and ensure consistency between the generated outputs and target data. Hong et al. \citep{hong2022source} propose a style-compensation transformation architecture guided by BN statistics stored in the source model and the reliable pseudo-labels generated from the target domain to generate source-like images. Rather than directly generating source-like images, certain studies propose aligning the feature prototypes or feature distribution of source data with those of the target domain. For example, Stan et al. \citep{stan2021unsupervised, stan2021privacy} propose a method that starts by generating a prototypical distribution in an embedding feature space, representing the source data using a Gaussian Mixture Model (GMM). They then perform source-free adaptation using the sliced Wasserstein distance to enforce distribution alignment between the source and target domains.

\textbf{Pseudo-label approach:} 
Considering the unreliable source model predictions on the target domain, certain studies \citep{chen2021source, xu2022denoising} address domain shift by effectively removing noise from unreliable target pseudo-labels. Chen et al. \citep{chen2021source} propose a prediction denoising method for cross-domain segmentation tasks. A crucial aspect of their approach involves pixel-wise denoising through uncertainty evaluation using Monte Carlo Dropout. This technique calculates the standard deviation of multiple stochastic outputs and ensures it remains below a manually-defined threshold. By filtering out the noisy pseudo-labels using this method, the quality of the pseudo-labels is improved, leading to more effective adaptation in the segmentation task. Similarly, Xu et al. \citep{xu2022denoising} propose an adaptive class-dependent threshold strategy as an initial denoising step to generate pseudo labels. They further introduce an uncertainty-rectified label soft correction method for fine denoising, which involves estimating the joint distribution matrix between the observed and latent labels. Recently, Zhou et al. \citep{zhou2023superpixel} propose a Superpixel-guided Class-level Denoised self-training framework (SCD) consisting of three components that work together to enhance the effectiveness of fundus image segmentation.

Few studies employ batch statistics from pre-trained source models to approximate the distribution of unavailable source data. This approximation is then used to mitigate the distribution discrepancy between the source and target domains during cross-domain adaptation. For instance, Liu et al. \citep{liu2021adapting, liu2023memory} suggest a method for adjusting the domain-specific low-order batch statistics, such as mean and variance, by using an exponential momentum decay approach. They also ensure the consistency of domain-shareable high-order batch statistics, such as scaling and shifting parameters. Additionally, they measure the transferability of each channel by evaluating its inter-domain divergence and make the assumption that channels with lower divergence have a greater impact on domain adaptation. Some works apply knowledge distillation in a self-supervised manner to transfer knowledge from source data to the target model. For example, Liu and Yuan \citep{liu2022source} introduce a self-supervised distillation approach for automatic polyp detection. They ensure the output consistency between weakly and strongly augmented polyp images, implicitly transferring source knowledge to the target model through a mean teacher strategy. Moreover, they propose a diversification flow paradigm to gradually reduce style sensitivity across different domains, thereby improving the model's resilience to style variations.

\textbf{Entropy minimization loss:} 
Bateson et al. \citep{bateson2020source, bateson2022source} utilize an entropy minimization loss to reduce the uncertainty of model predictions. Specifically, Bateson et al. \citep{bateson2020source} replace the conventional supervised loss used in the source domain with a direct minimization of label-free prediction entropy in the target domain. To avoid trivial solutions, they incorporate a regularization term into the entropy loss. This regularization is based on a class-ratio prior, which is derived from approximate anatomical knowledge. It is then integrated into the total loss function as a Kullback-Leibler (KL) divergence. To extend their previous work, the authors in \citep{bateson2022source} introduce a new loss function that addresses source-free adaptation. This novel approach incorporates an intriguing perspective based on mutual information and offers improved gradient dynamics compared to the loss function introduced in \citep{bateson2020source}.

To utilize diverse and rich information from multiple domains, Wang et al. \citep{wangmetateacher} bring together the challenges of multi-SFDA and semi-supervised learning and propose a framework called MetaTeacher based on a multi-teacher and one-student paradigm. It has three key components: (1) a learnable coordination scheme for adaptive domain adaptation of individual source models; (2) a mutual feedback mechanism between the target model and source models for more coherent learning; and (3) a semi-supervised bilevel optimization algorithm for reliable planning of the adaptation of source models and the learning of the target model. Detailed experiments using five chest X-ray image datasets demonstrate the effectiveness of the method.

\subsection{Test time adaptation (TTA)}

Test-time adaptation (TTA) focuses on adapting the model to unlabeled online data from the target domain without access to the source data \citep{song2023ecotta}. TTA requires a single test data \citep{10.1007/978-3-030-59710-8_43, he2021autoencoder} or a mini-batch of test data \citep{ma2022test} for model tuning. This tuning is often performed in an online manner without human supervision \citep{sun2020test}. TTA offers two primary advantages. Firstly, it eliminates the need for iterative training, resulting in enhanced computational efficiency. This allows the model to be readily deployed in real-time or online scenarios. Secondly, TTA does not rely on target training data, enabling the model to exhibit strong generalization capabilities across diverse target domains. 

Varsavsky et al. \citep{10.1007/978-3-030-59710-8_42} propose an evaluation framework where they perform test-time UDA on each subject separately. For the segmentation task, Bateson et al. \citep{10.1007/978-3-031-16440-8_70} perform single-subject TTA. During the inference stage on a test subject, they minimize the entropy of predictions and apply a class-ratio prior over batch normalization parameters. Additionally, they incorporate shape priors through penalty constraints to facilitate adaptation. He et al. \citep{10.1007/978-3-030-59710-8_43, he2021autoencoder} enhance a conventional deep network (task model) by incorporating two additional components: autoencoders and adaptors. This modification aims to provide the task model with the ability to handle unknown shifts in test domains effectively. For multi-center domain adaptation (DA), Karani et al. \citep{karani2021test} propose a method where the source segmentation model is trained by concatenating two sub-networks. The first sub-network is a relatively shallow image normalization CNN, followed by a deep CNN that performs segmentation on the normalized image. During the testing phase, the image normalization sub-network adjusts to each test image using an implicit prior guided by predicted segmentation labels. To model this implicit prior for plausible anatomical segmentation labels, they utilize a denoising autoencoder trained independently.

Existing test-time adaptation (TTA) methods suffer from a common limitation of using a fixed learning rate for all test samples. This approach is suboptimal because test data may arrive sequentially, resulting in frequent changes in the scale of distribution shift. To overcome this challenge, Yang et al. \citep{yang2022dltta} propose a novel method called DLTTA (Dynamic Learning Rate Adjustment for Test-Time Adaptation). DLTTA dynamically adjusts the amount of weight updates for each test image to account for variations in distribution shift. The method utilizes a memory bank-based estimation method to quantify the disparity of a specific test image. Using this estimated discrepancy, DLTTA incorporates a dynamic learning rate adjustment strategy to achieve an optimal level of adaptation for each test sample.

Some studies perform TTA \citep{pandey2020target, ma2022test} for classification tasks. For White Blood Cells (WBCs) classification taken from different imaging modalities, Pandey et al. \citep{pandey2020target} approach the problem of UDA by framing it as the task of finding the ``closest-clone" in the source domain for a given target image. This closest-clone serves as a proxy for the target image in the classifier trained on the source data. Authors demonstrate the existence of this clone given that an infinite number of data points can be sampled from the source distribution. To identify the clone, an optimization strategy over the latent space based on deep generative model is utilized. Ma et al. \citep{ma2022test} handles the label distribution shift by expanding the idea of balanced softmax loss to mimic multiple distributions where one class dominates other classes in order to learn representative classifiers with distribution calibration. Assess the technique for the two critical medical applications of COVID-19 severity prediction and liver fibrosis staging.

%%%%%%%%%%%%%%%%%%%%%%%%%%%%%%%%%%%%%%%%%%%%%%%%%%%%%%%%%%%%%%%%%%5

\section{Future research scope \label{sec_5}}

\subsection{Open/partial/universal-set domain adaptation}

This survey specifically focuses on close-set domain adaptation, where the label space of the source and target domains remains the same, i.e., $C_s = C_t$, where $C_s$ denotes the label set of the source domain and $C_t$ denotes the label set of the target domain. However, practical scenarios become more complex when category shifts occur across different domains. There are three types of non-close-set scenarios: (1) open-set DA problems where the source category label set is a subset of the target category label set ($C_s$ $\subset$ $C_t$), (2) partial-set DA where the source category label set is a superset of the target category label set ($C_s$ $\supset$ $C_t$), and (3) universal-set DA where no prior knowledge are
required on the label set relationship between domains \citep{you2019universal}.

To tackle the challenge of category shift in domain adaptation, various solutions have been proposed from different perspectives. These include approaches such as the distribution weighted combining rule \citep{xu2018deep}, the one-vs-all learning scheme \citep{saito2021ovanet}, the construction of out-of-distribution data \citep{kundu2020universal}, and the neighborhood clustering learning \citep{saito2020universal}. However, in the context of medical imaging, only a limited number of studies \citep{meng2020mutual, gao2022hierarchical, zhou2022delving} have specifically addressed the category shift problem in domain adaptation. Meng et al. \citep{meng2020mutual} propose end-to-end trainable Mutual Information-based Disentangled Networks (MIDNet) for learning generalized categorical
features to classify unseen categories in the target domain. Gao et al. \citep{gao2022hierarchical} propose a novel Hierarchical Feature Disentangling Network (HFDN) for universal domain adaptation (UniDA). Zhou et al. \citep{zhou2022delving} investigate the open source domain adaptation issue for fundus disease recognition by focusing on local features. Their proposed approach involves using collaborative regional clustering and alignment to identify common local feature patterns which are category-agnostic. Considering that addressing these different types of category shifts in medical imaging is crucial for ensuring the robustness, accuracy, and generalizability of models across different clinical settings and patient populations. We encourage further research to address this topic in the future.

\subsection{Continual/lifelong domain adaptation}

Continuous DA considers the adaptation problem with target data that is continuously changing, in contrast to standard DA which considers a specific target domain. Continual learning \citep{de2021continual} and lifelong learning \citep{parisi2019continual} are closely linked to the continuous adaption challenges as a potential solution to catastrophic forgetting \citep{wang2022continual}. There are two basic challenges with continuous UDA. First, there should be no catastrophic forgetting in the model, which means that it should continue to perform well on both the source domain and any previously encountered target domains. Second, the model should accurately adapt to new target domains without having access to the source or earlier target domains \citep{chen2023generative}. Taking inspiration from the field of continual/lifelong learning, recent advancements in continual domain adaptation have shown significant progress. These advancements include the exploration of gradient regularization, iterative neuron restoration, buffer sample mixture, and other techniques. However, in the domain of medical imaging, continual domain adaptation remains a relatively underdeveloped area, with only a few studies addressing this topic \citep{lenga2020continual, chen2023generative}. Lenga et al. \citep{lenga2020continual} look into methods used in the area of continuous learning for DA. Generative Appearance Replay for continuous Domain Adaptation (GarDA) \citep{chen2023generative} is a generative-replay based method that can incrementally adapt a segmentation model to new domains with unlabeled data. Given the dynamic and ever-changing nature of medical data, we believe continual domain adaptation is worth investigating for future work.

\subsection{Dynamic architectures}

The weights in a convolutional neural network (CNN) serve as feature detectors and typically remain fixed after being trained on source domains. However, this fixed nature of the weights can limit the CNN's ability to effectively represent data from unseen domains, resulting in poor generalization when the image statistics significantly differ in those domains. One potential solution is to explore dynamic architectures \citep{han2021dynamic}, where the network's weights are conditioned on the input \citep{jia2016dynamic}. This approach aims to make the network parameters dependent on the input while keeping the model size manageable for efficient computation. Dynamic architectures, such as dynamic filter networks \citep{jia2016dynamic} and conditional convolutions \citep{yang2019condconv}, have demonstrated effectiveness in various generic visual recognition tasks, including classification and segmentation. It would be intriguing to investigate whether these flexible architectures can be leveraged to address the challenge of domain shift in domain adaptation.

\subsection{Flexible target model design}
The utilization of automated approaches, such as neural architecture search (NAS) \citep{elsken2019neural, ren2021comprehensive}, is expected to enhance the learning performance of target models. NAS is a technique that automates the process of designing neural network architectures. Integrating NAS into UDA scenarios can help identify more suitable and efficient target models. However, it is important to carefully consider the balance between the search space and search cost of network parameters when applying NAS to UDA. Additionally, other hyperparameters used in NAS, such as the optimizer strategy and weight decay regularization, should be chosen and tuned appropriately to achieve optimal network performance \citep{ren2021comprehensive}.

\section{Conclusion \label{sec_6}}

Over the past few years, extensive research has been conducted on unsupervised domain adaptation in medical images, resulting in the development of numerous methods across various application areas. In this paper, we present a comprehensive review that encompasses recent advances in deep UDA for medical imaging. Specifically, we provide a detailed discussion of deep UDA methods in medical imaging and categorize them into six groups. Moreover, we further subdivide these methods based on the specific tasks they address, such as classification, segmentation, detection, medical image synthesis, and other related tasks. Additionally, we analyze the respective datasets utilized in different UDA studies to evaluate the divergence between the source and target domains. Finally, we investigate emerging areas within this research area and highlight several potential future directions for research and development. We believe that this survey paper will support researchers in gaining insights into the state-of-the-art in this area and contribute to further advancements in the domain.

%%%%%%%%%%%%%%%%%%%%%%%%%%%%%%%%%%%%%%%%%%%%%%%%%%%%%%%%%%%%%%%%%%
%%%%%%%%%%%%%%%%%%%

\bibliographystyle{elsarticle-num}
\bibliography{refs}
%\bibliographystyle{model2-names.bst}\biboptions{authoryear}
%\bibliographystyle{model2-names.bst}
%\bibliography{refs}

\end{document}

%% file: defs.tex
\def\x{{\bf x}}

\def\0{{\bf 0}}
\def\1{{\bf 1}}